\documentclass[aps,prb,showpacs,superscriptaddress,twocolumn,10pt,floatfix]{revtex4-1}
\usepackage{graphicx}
\usepackage{natbib}
\usepackage{amsmath}
\usepackage{epsfig}
\usepackage{physics}
\usepackage{multirow}
\usepackage{siunitx}
\usepackage{natbib}
\usepackage{indentfirst}
\usepackage{siunitx}
\usepackage{mathtools}
\usepackage{amssymb}
\usepackage{bbold}
\usepackage{xcolor}
\usepackage{bm}
\usepackage{eufrak}
\usepackage{epstopdf}
\usepackage{fancyhdr}

\begin{document}

\title{SU(4) spin waves in the $\nu=\pm1$ quantum Hall ferromagnet in graphene}

\author{Jonathan Atteia}
\email{jonathan.atteia@u-psud.fr}
\affiliation{Laboratoire de Physique des Solides, Universit\'e Paris Saclay, CNRS UMR 8502, F-91405 Orsay Cedex, France}

\author{Mark Oliver Goerbig}
\email{mark-oliver.goerbig@universite-paris-saclay.fr}
\affiliation{Laboratoire de Physique des Solides, Universit\'e Paris Saclay, CNRS UMR 8502, F-91405 Orsay Cedex, France}

\date{\today}
\begin{abstract}
We study generalized spin waves in graphene under a strong magnetic field when the Landau-level filling factor is $\nu=\pm 1$. In this case, the ground state is a particular SU(4) quantum Hall ferromagnet, in which not only the physical spin is fully polarized but also the pseudo-spin associated with the valley degree of freedom. The nature of the ground state and the spin-valley polarization depend on explicit symmetry breaking terms that are also reflected in the generalised spin-wave spectrum. In addition to pure spin waves, one encounters valley-pseudo-spin waves as well as more exotic entanglement waves that have a mixed spin-valley character. Most saliently, the SU(4) symmetry-breaking terms do not only yield gaps in the spectra, but under certain circumstances, namely in the case of residual ground-state symmetries, render the originally quadratic (in the wave vector) spin-wave dispersion linear.  
\end{abstract}

\maketitle

\section{Introduction}

Graphene, a one-atom-thick layer of carbon atoms arranged in a honeycomb lattice, is the prototype of a large class of two-dimensional materials such as transition metal dichalchogenoids\cite{Manzeli2017}, van der Waals heterostructures\cite{Geim2013} or twisted bilayers\cite{Lu2019} and multilayers\cite{Sinha2020} that present striking properties such as topological, correlated or superconducting phases. It is the paradigm of Dirac fermions in condensed matter since its dispersion is described by the Dirac-Weyl equation in two dimensions\cite{Novoselov2005,Cayssol2013}. These fermions come in two flavours with different chiralities, represented here by the valley index, which acts as an effective "pseudo-spin". 

Upon the application of a magnetic field $B$ perpendicular to the graphene plane, the relativistic character of the Dirac fermions is at the origin of an anomalous quantum Hall effect. While the effect is still a consequence of the quantization of the electrons' energy into highly degenerate Landau levels (LLs), the latter inherit from the $B=0$ system a twofold valley degeneracy, in addition to the spin degeneracy, such that the low-energy Hamiltonian is invariant under SU(4) spin-valley transformations. This SU(4) symmetry is furthermore respected to leading order by the Coulomb interaction between the electrons, which constitutes the dominant energy scale in partially filled LLs due to the flatness of the latter. If only some spin-valley branches of a specific LL are filled, all the electrons inside this LL choose to spontaneously break the SU(4) symmetry and to be polarized in a certain spin and pseudo-spin state. This marks the onset of \textit{SU(4) quantum Hall ferromagnetism}\cite{Nomura2006,Doretto2007,Goerbig2011,Young2012}.

The physics inside a LL is thus dominated by the Coulomb interaction $E_C=e^2/\varepsilon l_B=625\sqrt{B[T]}\si{K}/\varepsilon$, where $\varepsilon$ is the dielectric constant of the environment the graphene sheet is embedded into, and $l_B=\sqrt{\hbar/eB}$ is the magnetic length. However, at much smaller energies, explicit symmetry breaking terms become relevant, such as the Zeeman term, short-range electron-electron interactions, electron-phonon interactions or coupling to the substrate\cite{Alicea2006,Abanin2006,Sheng2007,Nomura2009,Kharitonov2012}. These symmetry-breaking terms, which happen to be all on the same order of magnitude, determine thus the spin-valley polarization of the ground state. At half-filling of the $n=0$ LL ($\nu=0$) several phases have been proposed such as a ferromagnetic (F), charge density wave (CDW), Kekulé distortion (KD) and canted anti-ferromagnetic (CAF) phase as a function of the symmetry breaking terms\cite{Herbut2007,Herbut2007a,Kharitonov2012,Duric2014}. Notice that there is experimental evidence for three of these phases\cite{Young2014,Li2019,Veyrat2020}, indicating that the nature of the SU(4) ferromagnetic ground state may be sample and/or substrate dependent. At quarter filling $\nu=\pm1$ -- the signs are related by particle-hole symmetry -- the phase diagram has been obtained by Lian \textit{et al.} \cite{Lian2017} using the same symmetry breaking terms as Kharitonov\cite{Kharitonov2012}, and one obtains similar phases as in the $\nu=0$ case.

Spin waves are the lowest energy excitations in a ferromagnet. They have been observed in a wide variety of materials\cite{Berger1996,Tsoi2000,Kajiwara2010a,An2014,Hamadeh2014,Collet2016a,Wimmer2019} and are promising platforms for spintronics\cite{Wolf2001,Chumak2015}. In a two-dimensional electron gas (2DEG) in GaAs/AlGaAs heterostructures at filling $\nu=1$, the first example of a \textit{quantum Hall} ferromagnet, the ground state consists of all spins pointing in the direction of the magnetic field, and the spin waves correspond simply to the precession of the spins around their ground state position. Generalized spin waves have also been extensively studied and observed in bilayer 2DEGs where the layer index plays the role of the pseudo-spin. When the distance $d$ is on the order of the magnetic length $l_B$, quantum Hall ferromagnetism of the layer pseudo-spin is observed and manifests itself in the form of a global phase coherence between electrons in the two layers\cite{MacDonald1990,Moon1995}. At $\nu=1$ (quarter filling of the $n=0$ LL), the ground state is an interlayer coherent state where each electron is in a superposition of the two layers, and the physical spin is fully polarized. This ground state can be viewed as a condensate of electron-hole pairs which then possesses a gapless, linearly dispersing superfluid mode\cite{Wen1992,Fertig1989,MacDonald2001,Eisenstein2004}. This mode was observed experimentally\cite{Spielman2001} using tunneling spectroscopy. Put differently, this superfluid mode is associated with a U(1) symmetry of the ground state that corresponds to the phase of the electron-hole superposition. At $\nu=2$ (half-filling of the $n=0$ LL), one is confronted with a frustrated situation: a complete spin polarization excludes a full pseudo-spin polarization, and \textit{vice versa}. Depending on the relative strength of the Zeeman and interlayer tunneling term, the ground state can thus be a spin ferromagnet, a spin-singlet or an intermediate phase with CAF order\cite{Yang1999,Demler1999}. The dispersion of the modes at $\nu=2$ are presented in Ref. [\onlinecite{Hama2012a}].  The peculiarity of the CAF phase is that it possesses a U(1) symmetry associated with the invariance under spin rotation around the $z$ axis. Such a symmetry implies also a gapless linearly dispersing mode which was observed experimentally by inelastic light scattering \cite{Pellegrini1998} and nuclear magnetic resonance\cite{Kumada2006,Kumada2007}.

In graphene, due the SU(4) spin-valley symmetry, one can have valley pseudo-spin waves in addition to spin waves, and what we call ``entanglement'' waves of mixed spin-valley character. Recent experiments\cite{Stepanov2018,Wei2018,Zhou2020,Assouline2021} have managed to electrically emit and detect spin waves\cite{Takei2016} using local gates. This is a highly promising result in the prospective of probing and controlling the spin degree of freedom in quantum-Hall systems. So far, the observed threshold for the emission of a spin wave is equal to the size of the Zeeman gap, a strong indication of the emission a pure spin wave. However, Ref. [\onlinecite{Wei2020}] has suggested a setup susceptible to generate valley waves at the edge located at the interface between two regions with filling factors $(\nu_1,\nu_2)=(+1,-1)$. The full dispersion relation of spin waves in graphene at $\nu=0$ has been studied in Refs. [\onlinecite{Lambert2013}] and [\onlinecite{DeNova2017}], while the low-energy dispersion and gaps of the KD and CAF state spin-waves was obtained using a non-linear sigma model in Ref. [\onlinecite{Wu2014a}], which showed the presence of gapless linearly dispersing modes in these two phases. Ref. [\onlinecite{Wei2020}] has studied the transmission of spin waves at a junction between regions with different filling factors.

Motivated by these recent experiments considering interfaces between regions at $\nu=1$, 0 and $-1$, we present in this paper a classification of the dispersion relations and the associated gaps in the graphene quantum Hall ferromagnet at $\nu=1(-1)$ when one sub-LL is empty (filled). We consider the spin waves in the four phases introduced in Ref. [\onlinecite{Lian2017}] with the addition of a ``valley Zeeman'' term. However, since this term does not modify substantially the phases but rather the location of their phase transitions, we consider only the dispersion in the phases of Ref. [\onlinecite{Lian2017}]. At $\nu=-1$, there are three Goldstone mode corresponding to flipping one electron from the filled sub-LL to each one of the three empty sub-LLs. In the simple phases such as KD or CDW, the three modes correspond to a pure spin wave, a pseudo-spin wave and an entanglement wave. We derive a non-linear sigma model valid at long wave lengths generalized to the CP$^3$ coset space corresponding to the space of broken symmetries. In the absence of explicit symmetry-breaking terms at low energies, all the dispersions are gapless and quadratic in the wave vector, corresponding thus to true Goldstone modes. In the presence of the symmetry breaking terms, some modes acquire a gap, while others remain gapless but acquire a linear dispersion relation until a certain momentum at which they recover their quadratic dispersion at higher momentum. We find that this behavior originates from a residual symmetry of the ground state. We also find that at several high-symmetry points in the phase diagram, some originally gapped modes become gapless.

The paper is organized as follows. In Sec. \ref{sec:QHFM}, we present the phase diagram originally introduced in Ref. [\onlinecite{Lian2017}] using a different labelling for the phases and also discuss the introduction of a valley Zeeman term. In Sec. \ref{sec:NLSM}, we present our non-linear sigma model using a Lagrangian formalism, while in Sec. \ref{sec:Disp}, we present our results for the dispersion relation in the different regions of the phase diagram. In the conclusion section, we present a summary of the various spin waves one encounters in each phase, in view of their dispersion, i.e. whether they are quadratic and gapped or linear and gapless.

\section{QHFM ground state}

\label{sec:QHFM}

In a single particle picture, flat Landau levels (LLs) are formed in graphene under a magnetic field with energies $E_{\lambda n}=\lambda\hbar\omega_c\sqrt{n}$ where $\lambda=\pm$ is the band index, $n$ is the LL index, $\omega_c=\sqrt{2}v/l_B$ is the cyclotron energy, and $v$ is the Fermi velocity of graphene. For a sufficiently strong magnetic field, the low-energy physics of a quantum Hall ferromagnet in the $n=0$ LL is dominated by the Coulomb interaction 
\begin{align}
\hat{V}_C=\frac{1}{2}\sum_{\mathbf{q}\neq 0}v(\mathbf{q})\bar{\rho}(\mathbf{q})\bar{\rho}(-\mathbf{q}),
\end{align}
in terms of the Coulomb potential multiplied by the lowest Landau level (LLL) form factor,
\begin{align}
v(\mathbf{q})=\frac{1}{\mathcal{A}}\frac{2\pi e^2}{\varepsilon|\mathbf{q}|}|\mathcal{F}_0(\mathbf{q})|^2,
\end{align}
where $\mathcal{A}$ is the area of the sample and $\mathcal{F}_0(\mathbf{q})$ is the form factor of the LLL (see eg. Ref. [\onlinecite{Goerbig2011}]). Furthermore, $\bar{\rho}(\mathbf{q})$ represents the density operator in momentum space projected into the LLL. This Hamiltonian is approximately SU(4) invariant under spin-valley rotations. The exchange terms favors a completely antisymmetric orbital wavefunction to minimize the Coulomb replusion, which then favors a completely symmetric spin-valley spinor. At filling $\nu=-1$,
there is thus one electron per orbital site and the uniform ground state is described by the Slater determinant 
\begin{align}
	|\psi_0\rangle=\prod_m\left(\sum_\mu F_\mu c^\dagger_{m,\mu}\right)|0\rangle
	\label{eq:ground_state}
\end{align}
where $\mu=\{\sigma,\xi\}$ runs over the spin ($\sigma\in\{\uparrow,\downarrow\}$) and valley ($\xi\in\{K,K'\}$) indices, $m$ is the Landau site index and $F$ is a normalized four-component spinor which describes the QHFM ground state. 

\subsection{Parametrization of the spinor}

The Coulomb Hamiltonian is SU(4) symmetric, while the broken symmetry ground state is invariant under SU(3)$\otimes$U(1) rotations corresponding to rotations between the three empty sub-LL and the relative phase between the empty and filled sub-LL. The coset space is thus $CP^3=U(4)/U(3)\otimes U(1)$ which has 6 real dimensions\cite{Yang2006}. A general spinor describing the broken symmetry ground state is thus parametrized by 6 angles. In order to describe the spinor $F$, we express it as a Schmidt decomposition in the basis $\{|K\uparrow\rangle,|K\downarrow\rangle,|K'\uparrow\rangle,|K'\downarrow\rangle\}$ as\cite{Doucot2008,Lian2017}
\begin{align}
|F\rangle=\cos\frac{\alpha}{2}|\mathbf{n}\rangle|\mathbf{s}\rangle+e^{i\beta}\sin\frac{\alpha}{2}|-\mathbf{n}\rangle|-\mathbf{s}\rangle,
\label{eq:param}
\end{align}
where $|\mathbf{n}\rangle|\mathbf{s}\rangle=|\mathbf{n}\rangle\otimes|\mathbf{s}\rangle$ is the tensor product of the spinors 
\begin{align}
|\mathbf{n}\rangle&=\begin{pmatrix}
\cos\frac{\theta_P}{2} \\ \sin\frac{\theta_P}{2}e^{i\varphi_P}
\end{pmatrix}, 
\label{eq:param1} \\
|\mathbf{s}\rangle&=\begin{pmatrix}
\cos\frac{\theta_S}{2} \\ \sin\frac{\theta_S}{2}e^{i\varphi_S}
\end{pmatrix}, \label{eq:param2} 
\end{align}
acting in valley and spin spaces respectively. We have $\bm{\sigma}\cdot\mathbf{s}|\pm\mathbf{s}\rangle=\pm|\pm\mathbf{s}\rangle$ and $\bm{\tau}\cdot\mathbf{n}|\pm\mathbf{n}\rangle=\pm|\pm\mathbf{n}\rangle$, where
\begin{align}
\mathbf{s},\mathbf{n}=\begin{pmatrix}
\sin\theta_{S,P}\cos\varphi_{S,P}\\
\sin\theta_{S,P}\sin\varphi_{S,P} \\
\cos\theta_{S,P}
\end{pmatrix}
\end{align}
are the unit vectors on the spin and pseudo-spin Bloch spheres, respectively, with $\theta_S,\theta_P\in[0,\pi]$ and $\varphi_S,\varphi_P\in[0,2\pi]$. The angles $\alpha\in[0,\pi]$ and $\beta_1\in[0,2\pi]$ are the angles of the "entanglement" Bloch sphere of the particle\cite{Doucot2008}.  The spinors $|-\mathbf{s}\rangle$ and $|-\mathbf{n}\rangle$ are obtained from $|\mathbf{s}\rangle$ and $|\mathbf{n}\rangle$
by the replacement $\theta\rightarrow\pi-\theta$ and $\varphi\rightarrow\varphi+\pi$ such that we have $\langle \mathbf{s}|- \mathbf{s}\rangle = \langle \mathbf{n}|- \mathbf{n}\rangle=0$.

When $\theta_P=0(\pi)$, the vector $\mathbf{n}$ lies at the north (south) pole of the pseudo-spin Bloch sphere corresponding to a polarization in valley $K(K')$. Analogously, for $\theta_S=0(\pi)$, the vector $\mathbf{n}$ lies at the north (south) pole of the spin Bloch sphere corresponding to spin up (down) polarization. Finally, this parametrization includes the possibility of ``entanglement'' between the spin and the pseudo-spin. In fact, this decomposition of the spinors does not correspond to real entanglement between two particles because here it is the spin and pseudo-spin of the \textit{same} particle which is ``entangled'', and the Schmidt decomposition can be viewed as a decomposition of SU(4) spinors in the basis of SU(2)$\otimes$SU(2) spinors. Because of this reminiscence and the relevance of the spin and pseudospin magnetizations in experimental measurements, we will refer loosely to the angle $\alpha$ as entanglement angle for simplicity.

\subsection{Symmetry breaking terms}

Inspired by earlier works\cite{Kharitonov2012,Nomura2009,Lian2017,Atteia2021} that focus on short-range electron-electron\cite{Alicea2006} and electron-phonon\cite{Kharitonov2012} interactions at the lattice scale, we consider the local anisotropic Hamiltonian 
\begin{align}
H_A=\frac{1}{2}&\int d^2r\left\{U_\perp[P_x^2(\mathbf{r})+P_y^2(\mathbf{r})]+U_zP_z^2(\mathbf{r})\right\}\nonumber \\
-&\int d^2r\left\{\Delta_ZS_z(\mathbf{r})+\Delta_P P_z(\mathbf{r})\right\},
\label{eq:SB}
\end{align}
where 
\begin{align}
\mathbf{P}(\mathbf{r})&=\Psi^\dagger(\mathbf{r})(\sigma_0\otimes\bm{\tau})\Psi(\mathbf{r}),\\
\mathbf{S}(\mathbf{r})&=\Psi^\dagger(\mathbf{r})(\bm{\sigma}\otimes\tau_0)\Psi(\mathbf{r})
\end{align}
are the local spin and pseudo-spin densities, respectively, in terms of the vectors 
$\bm{\sigma}$ and $\bm{\tau}$ of Pauli matrices vectors acting in spin and pseudo-spin spaces, respectively, while $\sigma_0$ and $\tau_0$ are the identity matrices. In the following, we neglect the identity and consider $\bm{\sigma}\equiv\bm{\sigma}\otimes\tau_0$ and $\bm{\tau}\equiv\sigma_0\otimes\bm{\tau}$. The potentials $U_\perp$ and $U_z$ correspond to local interactions that act when two electrons are at the same position, and they act only in valley space thus favoring in-plane or out-of-plane pseudo-spin polarizations. The relative values of $\Delta_Z$, $\Delta_P$, $U_z$ and $U_\perp$ determine thus the spin or pseudo-spin polarization of the ground state.

The first term in Eq. (\ref{eq:SB}) represents the electrons' interaction with "frozen" in-plane phonons\cite{Nomura2009} and is estimated to be of the order of $U_\perp \sim 2.0B[(T)]K$. This term creates a Kekul\'e-like distortion. The term $U_z$ originates from short-range Hubbard type interactions\cite{Alicea2006} and intervalley scattering which originate from the SU(4) symmetry breaking the in Coulomb interaction\cite{Goerbig2006}. Out-of-plane phonons also contribute to $U_z$ and is estimated to be of the order of $\sim 0.5B[(T)]K$. The Zeeman coupling $\Delta_Z=g\mu_BB$ is of the order of $\sim 1.2B[(T)]K$. Finally, $\Delta_P$ corresponds to a staggered potential on the A and B sublattice which generates a mass term in the Dirac equation and can be generated by the interaction with a substrate, eg hexagonal Boron-Nitride (hBN)\cite{Hunt2013,Amet2013}. Due to the locking of the sublattice and valley indices in the $n=0$ LL, this term is analogous to a Zeeman term acting in pseudo-spin space, we thereby dub it "valley Zeeman" term. This terms favors a polarization in one valley and thus on one sublattice. The energies $U_\perp$ and $U_z$ are proportional to the perpendicular magnetic field\cite{Li2019} while $\Delta_z$ is proportional to the total magnetic field. Moreover, $\Delta_P$ is an intrinsic effect and thus independent of the magnetic field. Notice that these energy scales are all on the same order of magnitude and are likely to be strongly sample-dependent. We thus consider them, here, as tunable parameters that determine the phase diagram of the QHFM ground states as well as that of the skyrmions formed on top of these states. 

Applying the Hartree-Fock approximation, the energy of the anisotropic energy $E_A=\langle F|H_A|F\rangle$ can be expressed as\cite{Lian2017}
\begin{align}
E_A[F]=\frac{N_\phi}{2}\left[u_\perp\left(M_{P_x}^2+M_{P_y}^2\right)+u_zM_{P_z}^2 \right]\\
-N_\phi\left[\Delta_ZM_{S_z}+\Delta_PM_{P_Z}\right],
\label{eq:aniso_GS}
\end{align}
where $N_\phi=A/(2\pi l_B^2)$ is the number of flux quanta threading the area $A$ of the sample and
\begin{align}
\mathbf{M_{P}}&=\langle F|\bm{\tau}|F\rangle=\mathbf{n}\cos\alpha \\
\mathbf{M_{S}}&=\langle F|\bm{\sigma}|F\rangle=\mathbf{s}\cos\alpha
\label{eq:spin_mag_GS} 
\end{align}
are the spin and pseudo-spin magnetization respectively. The parameters $u_{\perp,z}$ are obtained as
\begin{align}
	u_{\perp,z}=\mathcal{V}^H_{\perp,z}-\mathcal{V}^F_{\perp,z}
\end{align}
where $\mathcal{V}^H_{\perp,z}$ and $\mathcal{V}^F_{\perp,z}$ are the Hartree and Fock potentials, respectively, associated with the potentials $U_{\perp,z}$. For a $\delta(\mathbf{r})$ interaction, at $\nu=\pm1$, the Hartree and Fock potentials are identical and thus cancel each other\cite{Lian2017}. We thus postulate a slightly non-local interaction.

As a function of the angles, we obtain the expression
\begin{align}
E_A[F]=&N_\phi\left[\frac{1}{2}\cos^2\alpha(u_\perp\sin^2\theta_P+u_z\cos^2\theta_P)\right.\nonumber\\
&\left.-\Delta_P\cos\alpha\cos\theta_S-\Delta_Z\cos\alpha\cos\theta_S\right].
\label{eq:aniso_angles}
\end{align}
The phase diagram is obtained by minimizing Eq. (\ref{eq:aniso_angles}). We first consider the phase diagram without the valley Zeeman term $\Delta_P$ in Sec. \ref{sec:phase_diag_without}, while we show its effect in Sec. \ref{sec:phase_diag_with}.

\subsection{Phase Diagram without valley Zeeman term}

\label{sec:phase_diag_without}

The phase diagram of the QHFM at $\nu=\pm1$ without the valley Zeeman term was calculated by Lian \textit{et al}\cite{Lian2017}. Here, we briefly review the different phases in order to discuss the spin waves associated with each ground state.  There is a $\mathbb{Z}_2$ redundancy in the parametrization of the spinors (see appendix of Ref. [\onlinecite{Lian2017}]) such that without loss of generality we can assume $\alpha\in\left[0,\pi/2\right]$.  Using this fact we can see that the anisotropic energy is minimized for $\cos\theta_S=1$ everywhere.

\begin{figure}[t]
	\begin{center}
		\includegraphics[width=8cm]{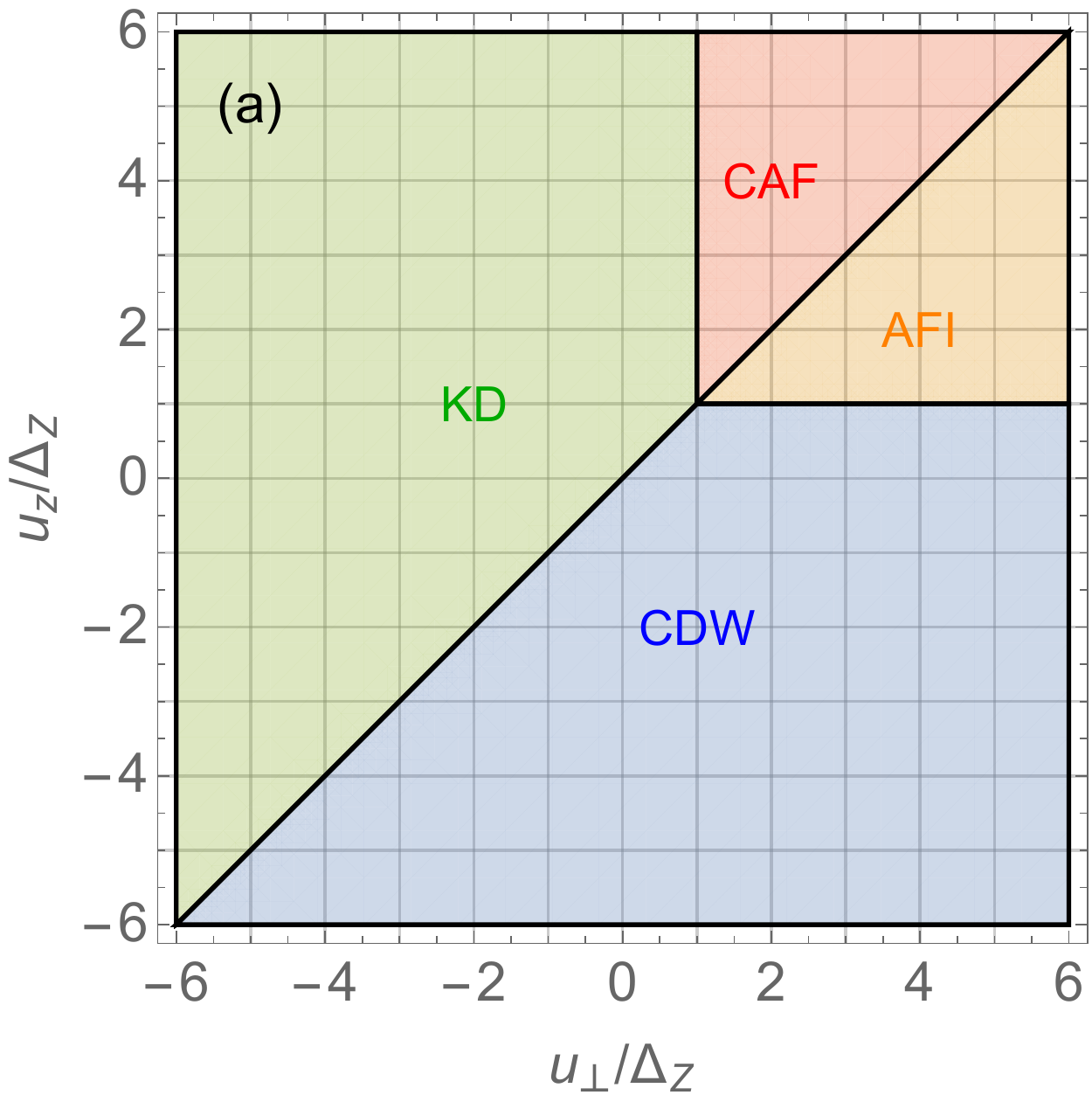} \\
		\includegraphics[width=4cm]{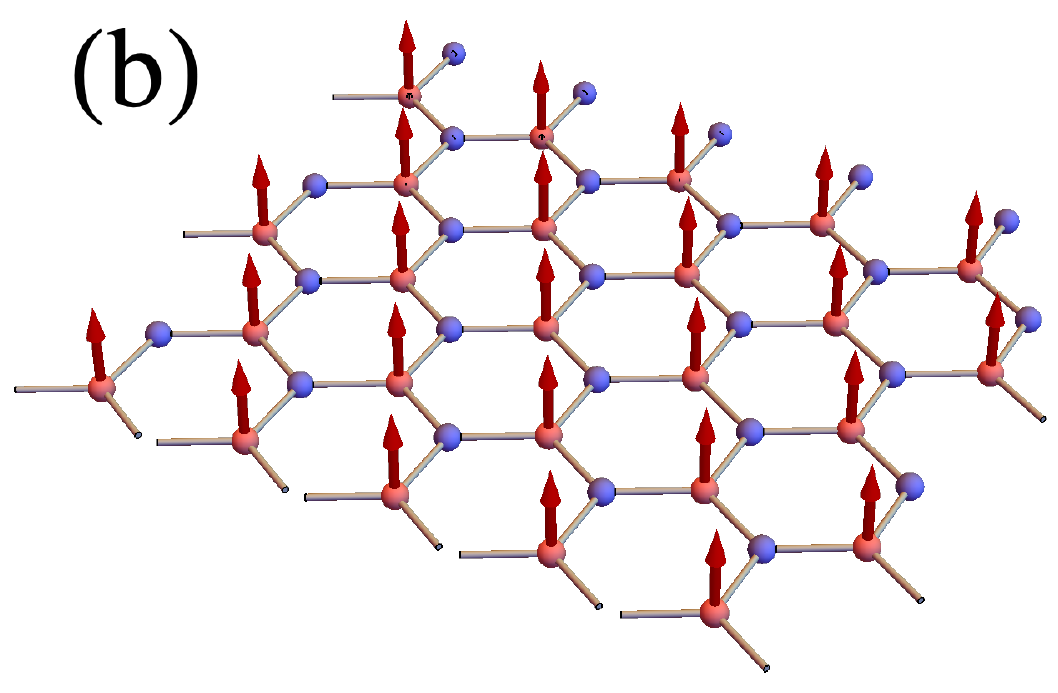}
		\includegraphics[width=4cm]{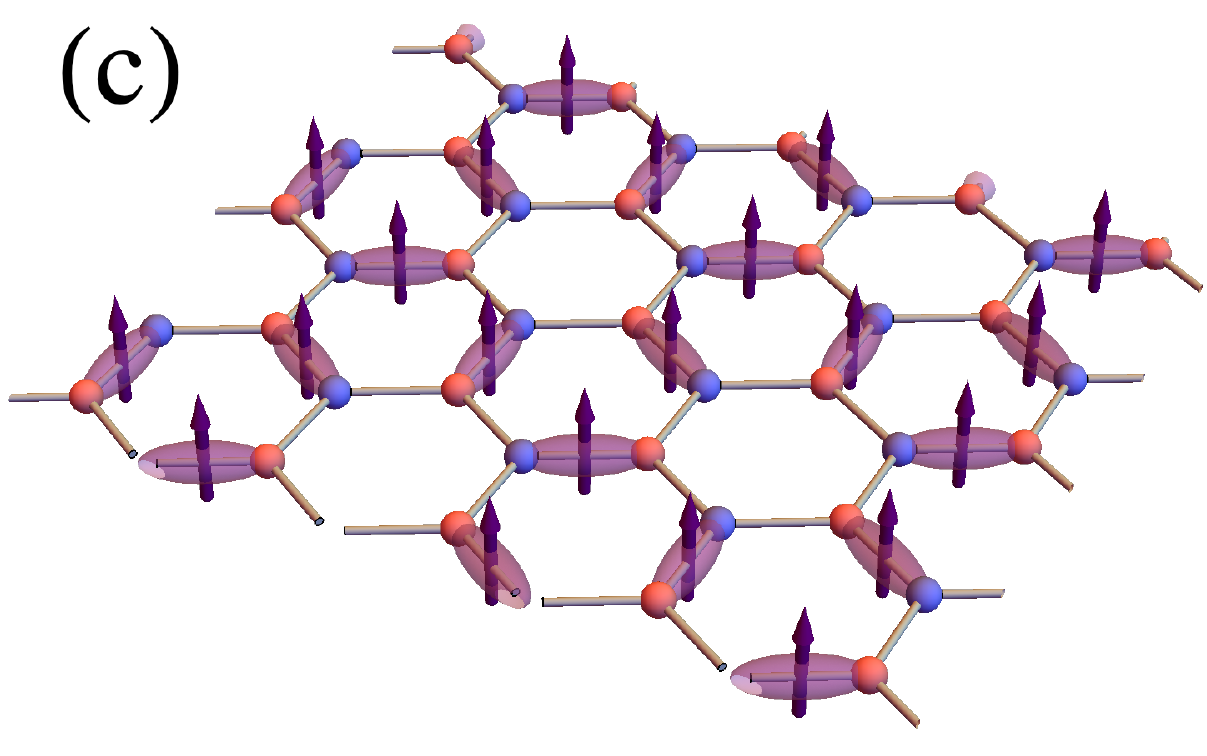} \\
		\includegraphics[width=4cm]{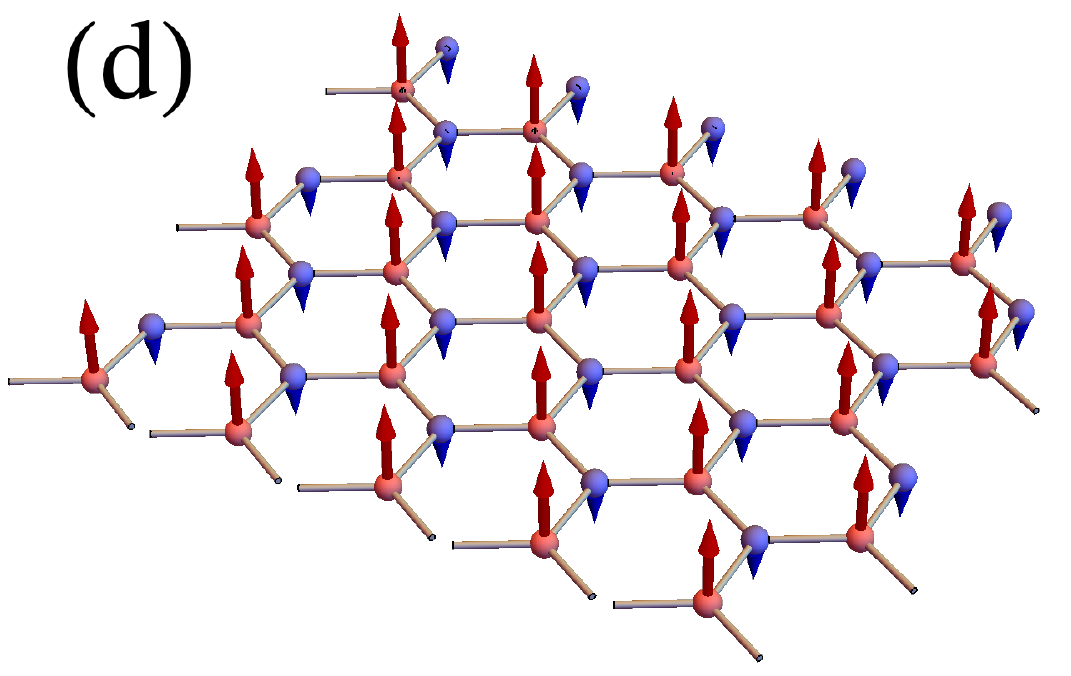}
		\includegraphics[width=4cm]{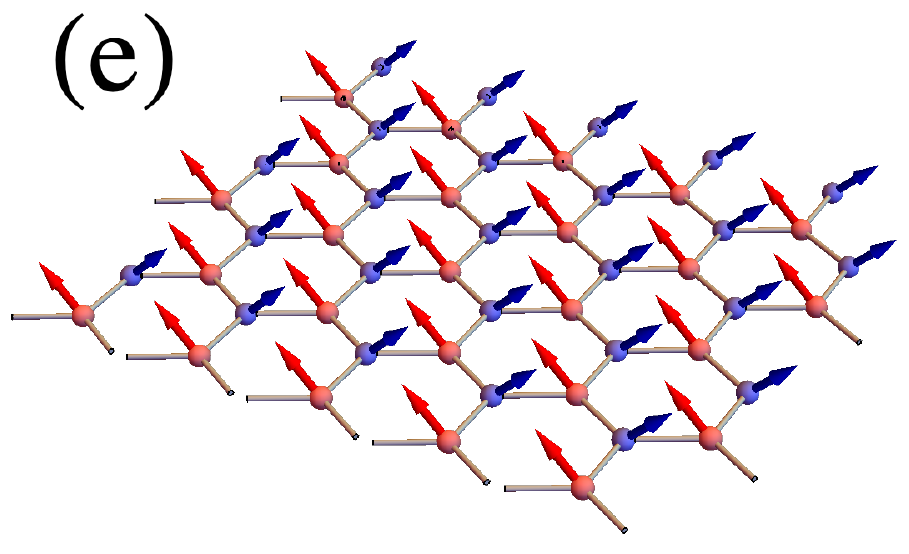}
		\caption{(a) Phase diagram of the QHFM ground state composed of four phase : charge density wave (CDW), Kekulé distortion (KD), anti-ferrimagnetic (AFI) and canted anti-ferromagnetic (CAF). (b)-(e) Spin magnetization on the A and B sublattices of the different phase. (b) CDW (c) KD, (d) AFI and (e) CAF.}
		\label{fig:phaseDiag}
	\end{center}
\end{figure}

Minimizing Eq. (\ref{eq:aniso_angles}), we find the four phases shown in Fig. (\ref{fig:phaseDiag}) which can be separated in two types : for $u_\perp>u_z$, an easy-axis pseudo-spin polarization is favored, which is the case of the charge density wave (CDW) and anti-ferrimagnetic (AFI) phases, while for $u_z>u_\perp$, an easy-plane polarization is favored, namely, the Kekulé distortion (KD) and canted anti-ferromagnetic (CAF) phase. In addition to that, the phases can present entanglement ($\alpha\neq0$) or not ($\alpha=\{0,\pi\}$). The CDW and KD phases are not entangled and they have maximal spin and pseudo-spin magnetizations, they are thereby ferromagnetic phases. The AFI and CAF phases are entangled, such that their spin and pseudo-spin magnetizations are reduced. These phases are realized in the regions of positive $u_\perp$ and $u_z$ because entanglement allows to reduce the pseudo-spin magnetization thus making a compromise between the spin and pseudo-spin magnetizations. In the limit of vanishing Zeeman term (compared to $u_\perp$ and $u_z$), these two phases are maximally entangled become both anti-ferromagnetic. We mention that, as opposed to the $\nu=0$ case, at $\nu=\pm1$, the spin \textit{and} pseudo-spin can be maximal at the same time. Thus the CDW and KD phases are pseudo-spin polarized and spin ferromagnetic, whereas at $\nu=0$, the phases can be either be spin polarized and pseudo-spin unpolarized (F), pseudo-spin polarized and spin unpolarized (KD and CDW) or entangled (CAF). Notice that in Ref. [\onlinecite{Lian2017}], these phases were named after their valley pseudo-spin magnetization : the CDW (AFI) phases are associated with an unentangled (entangled) easy-axis pseudo-spin order, while the KD (CAF) comes along with an unentangled (entangled) easy-plane pseudo-spin magnetization.

In order to characterize the different phases, we focus on experimentally measurable quantities such as the spin magnetization and electronic density on the A and B sublattices
\begin{align}
\rho_{A,B}&=\frac{1}{2}\langle F|(\tau_0\pm\tau_z)|F\rangle, \\
\mathbf{M_S}_{A,B}&=\frac{1}{2}\langle F|\bm{\sigma}(\tau_0\pm\tau_z)|F\rangle,
\end{align}
respectively.
 
The spinor of the CDW phase is
\begin{align}
	|F\rangle=|\mathbf{n}_z\rangle|\mathbf{s}_z\rangle,
\end{align}
where $\mathbf{n}_z=(1,0)^T$ and $\mathbf{s}_z=(1,0)^T$ correspond to a spin and pseudo-spin both polarized at the north of their respective Bloch spheres, such that the electrons have spin-up and are polarized in valley $K$ or $K'$ corresponding thus to a ferromagnetic phase restricted to a single sublattice. The sublattice polarization is given by $\rho_A=1$ and $\rho_B=0$ or $\rho_A=0$ and $\rho_B=1$ and there is thus a spontaneous $\mathbb{Z}_2$ sublattice symmetry breaking. The spin magnetizations on sublattices A and B are $\mathbf{M_{S_A}}=\mathbf{s}_z$ and $\mathbf{M_{S_B}}=0$.

The spinor of the KD phase is given by 
\begin{align}
|F\rangle=|\mathbf{n}_\perp\rangle|\mathbf{s}_z\rangle,
\end{align}
where $|\mathbf{n}_\perp\rangle=\frac{1}{\sqrt{2}}(1,e^{i\varphi})^T$ points to a position at the equator of the pseudo-spin Bloch sphere and corresponds thus to a superposition of the two valleys. The angle $\varphi$ corresponds to the orientation of the pseudo-spin magnetization in the $xy$ plane. There is thus a residual $U(1)$ symmetry corresponding to the angle $\varphi$. Both sublattices are equally populated such that $\rho_A=\rho_B=1/2$ and
$\mathbf{M_{S_A}}=\mathbf{M_{S_B}}=\frac{1}{2}\mathbf{s}_z$.

The spinor of the AFI phase has the expression
\begin{align}
|F\rangle&=\cos\frac{\alpha_1}{2}|\mathbf{n}_z\rangle|\mathbf{s}_z\rangle+e^{i\beta}\sin\frac{\alpha_1}{2}|-\mathbf{n}_z\rangle|-\mathbf{s}_z\rangle ,
\end{align}
with
\begin{align}
\cos\alpha_1=\frac{\Delta_Z}{u_z}.
\label{eq:alphaAFI}
\end{align}
This phase corresponds thus to an entangled phase which in turn reduces the amplitude of the spin magnetization in order to minimize the anisotropic energy. The spin magnetization on the A and B sublattices are $\mathbf{M_{S_A}}=\frac{1}{2}(1+\cos\alpha_1)\mathbf{s}_z$ and $\mathbf{M_{S_B}}=\frac{1}{2}(-1+\cos\alpha_1)\mathbf{s}_z$ such that the spin magnetization on each sublattice points along the $z$ direction but there is an imbalance between the spin magnetization in sublattices A and B. For $u_z=\Delta_Z$ ($\alpha_1=0$), namely at the CDW-AFI transition, we recover the CDW phase, while for $u_z\gg\Delta_Z$ ($\alpha_1\rightarrow\pi/2$), we have a maximally entangled phase with $\mathbf{M_{S_A}}=-\mathbf{M_{S_B}}=\frac{1}{2}\mathbf{s}_z$ which is anti-ferromagnetic, as we would expect in the limit of a vanishing Zeeman effect.

The spinor of the CAF phase has the expression
\begin{align}
|F\rangle&=\cos\frac{\alpha_2}{2}|\mathbf{n}_\perp\rangle|\mathbf{s}_z\rangle+e^{i\beta}\sin\frac{\alpha_2}{2}|-\mathbf{n}_\perp\rangle|-\mathbf{s}_z\rangle,
\end{align}
with
\begin{align}
\cos\alpha_2=\frac{\Delta_Z}{u_\perp}.
\label{eq:alphaCAF}
\end{align}
This phase has its pseudo-spin polarized in the $xy$ plane of the Bloch sphere and presents entanglement analogously to the AFI phase. Both sublattices are populated equally $\rho_A=\rho_B=1/2$. The spin magnetization on the A and B sublattices forms a canted anti-ferromagnetic pattern with $\mathbf{M_{S_{A,B}}}=(\pm\sin\alpha_2\cos(\beta-\varphi),\pm\sin\alpha_2\sin(\beta-\varphi),\cos\alpha_2)$ such that the $z$ component of the magnetization is identical on both sublattices, but there is a canting of the spin in the $xy$ plane with opposite orientation on the sublattices. At the transition with the KD phase ($\Delta_Z=u_\perp\rightarrow\alpha_2=0$), we recover a ferromagnetic phase with equal weight on the K and K' valleys, while in the fully entangled limit ($u_\perp\gg\Delta_z\rightarrow\alpha_2=\pi/2$), we obtain an anti-ferromagnetic phase with spins pointing in the $xy$ plane.

\subsection{Phase diagram with valley Zeeman}

\label{sec:phase_diag_with}

\begin{figure}[t]
	\begin{center}
		\includegraphics[width=8cm]{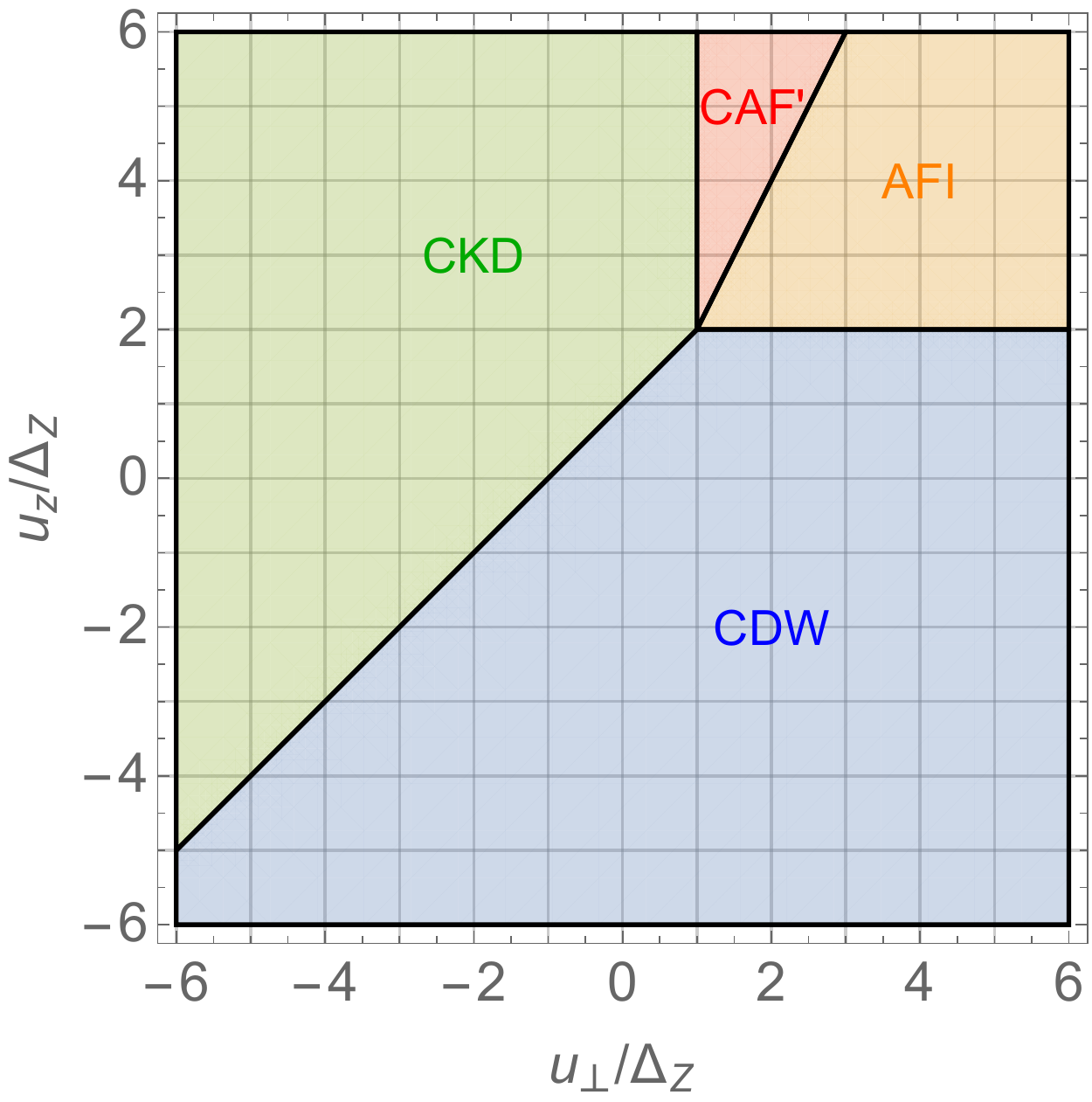}
		\caption{(a)Phase diagram of the QHFM ground state with the valley Zeeman term $\Delta_P$ such that $\Delta_P=\Delta_Z$. The KD and CAF phases are modified compared to the case without the valley Zeeman term and are turned to a canted KD phase (CDW) and a different CAF phase (CAF').}
		\label{fig:phaseDiagwith}
	\end{center}
\end{figure}

Experimentally, graphene is generally placed on top of a substrate. In the case of hBN, a potential difference is generated between the A and B sites of graphene and yields a valley-dependent potential due to the valley-sublattice equivalence in the LLL of graphene. Such a term favors a polarization on one sublattice and thus in one valley, analougously to a Zeeman term in valley space. The evolution of the phase diagram in the presence of the valley Zeeman term is shown in Fig. \ref{fig:phaseDiagwith}. The phases CDW and AFI are not modified by the valley Zeeman term because their pseudo-spin is already polarized in one valley. However, the presence of the valley Zeeman breaks the $Z_2$ symmetry between the two valleys by favoring one valley corresponding to the sublattice with smallest on-site potential. However, the KD and CAF phases are modified such that their pseudo-spin polarization is now canted towards the north pole of the Bloch sphere (or the south pole if the staggered potential is reversed). The KD phase becomes a canted KD phase with spinor
\begin{align}
	|F\rangle=|\mathbf{n}\rangle|\mathbf{s}_z\rangle,
\end{align}
with
\begin{align}
	\cos\theta_P=\frac{\Delta_P}{(u_z-u_\perp)}.
\end{align}
There is thus a continuous phase transition between the CDW and CKD phase transition located at  $u_z-u_\perp=\Delta_P$, where the pseudo-spin is progressively canted relative to the $z$ direction. For $u_z-u_\perp\gg\Delta_P$, we recover the KD phase. The CDW occupied thus a larger portion of the phase diagram compared to the $\Delta_P=0$ case (see Fig. \ref{fig:phaseDiag}).

The transition between the CDW and AFI phase is also modified because the cost to entangle the easy-axis phase implies a non-zero weight on the valley $K'$. Thereby, the transition occurs at $u_z=(\Delta_P+\Delta_Z)$ and the entanglement angle in the AFI phase $\alpha_1$ is now given by
\begin{align}
\cos\alpha_1=\frac{\Delta_Z+\Delta_P}{u_z}.
\end{align}

Finally, the CAF phase is also modified into a different CAF phase such that the spinor reads
\begin{align}
|F\rangle&=\cos\frac{\alpha_2}{2}|\mathbf{n}\rangle|\mathbf{s}_z\rangle+e^{i\beta}\sin\frac{\alpha_2}{2}|-\mathbf{n}\rangle|-\mathbf{s}_z\rangle,
\end{align}
where
\begin{align}
	\cos\alpha_2=\frac{\Delta_Z}{u_\perp}\quad\text{and}\quad\cos\theta_P=\frac{\Delta_Z}{\Delta_P}\frac{u_\perp}{(u_z-u_\perp)}.
\end{align}
Once again, the AFI phase is favored in a larger part of the phase diagram and the transition between the AFI and CAF phases is located at $u_z=u_\perp(\Delta_P/\Delta_Z+1)$. The four phase transitions meet at the point $(u_\perp,u_z)=(\Delta_Z,\Delta_Z+\Delta_P)$.

\section{Non-linear sigma model}

\label{sec:NLSM}

In order to find the dispersion relations of the Goldstone modes, we derive an effective Lagrangian which describes the low-energy (long-wavelength) excitations of the ground state. In the SU(4) invariant limit (in the absence of symmetry breaking terms), this Lagrangian consists of a non-linear sigma model describing the fields associated with the broken symmetries. The collective modes of this Lagrangian are the different Goldstone modes. In the presence of the symmetry breaking terms, the Goldstone modes acquire a mass gap.

\subsection{Broken symmetries and their generators}

\begin{figure}[t]
	\begin{center}
		\includegraphics[width=8cm]{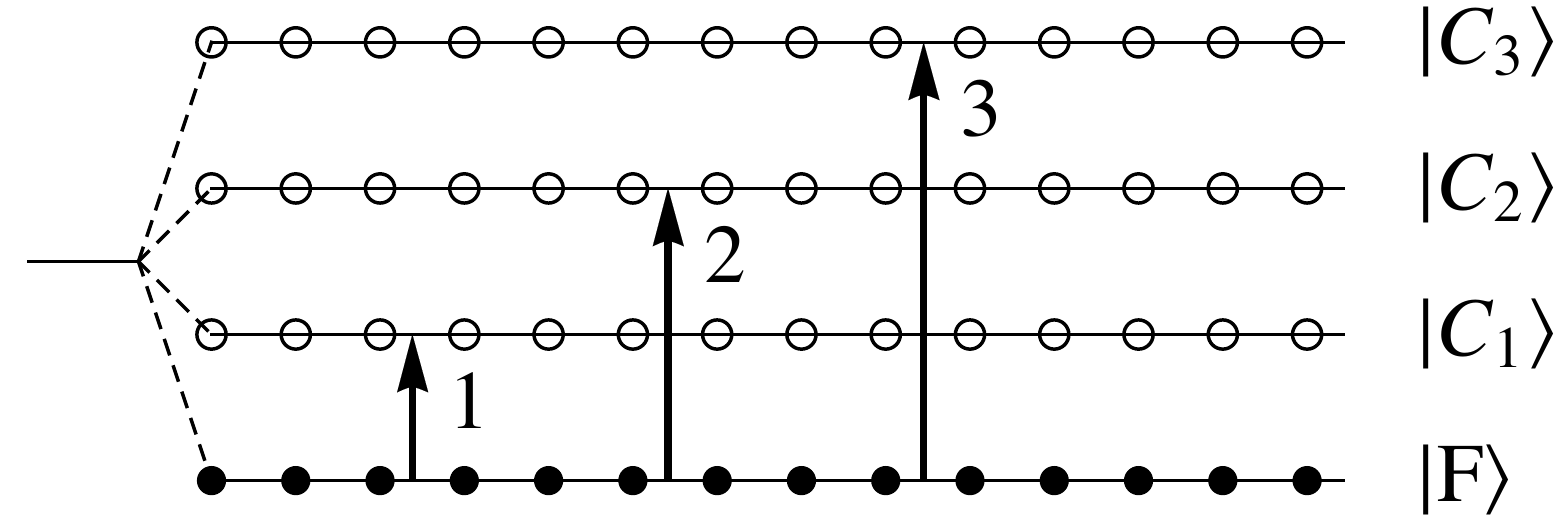} 
		\caption{Four sub-LLs of the $n=0$ LL and the three associated spin wave modes corresponding to the mixing of the filled sub-LL described by the spinor $|F\rangle$ with each of the three empty sub-LLs described by the spinors $|C_i\rangle$.}
		\label{fig:LLmodes}
	\end{center}
\end{figure}

At filling factor $\nu=\pm1$, the spontaneous symmetry breaking mechanism corresponds to filling one sub-LL out of the four with any SU(4) spin-valley orientation (in the absence of symmetry breaking term). Explicitely, this symmetry breaking mechanism corresponds to
\begin{align}
SU(4)\rightarrow SU(3)\otimes U(1),
\end{align}
where SU(4) is the original symmetry of the Hamiltonian which in composed of 15 generators and SU(3)$\otimes$U(1) is the residual symmetry the ground state which is invariant under tranformations that mixes the 3 empty sublevels corresponding to 8 generators times the relative U(1) phase between the empty and the occupied sub-LLs. According to Refs. [\onlinecite{Arovas1999}] and [\onlinecite{Yang2006}], there are thus $15-8-1=6$ generators associated with the broken symmetries. For simplicity, we label these generators "broken generators". The corresponding coset space of the non-linear sigma model is the complex projective space $CP^3=U(4)/[U(3)\otimes U(1)]$ which has six dimensions\cite{Yang2006}.

In order to find an explicit expression for the broken generators, we consider for simplicity the CDW ground state $|F\rangle=|\mathbf{n}_z\rangle|\mathbf{s}_z\rangle=|K\uparrow\rangle$ to be the filled sub-LL in the basis $\mathcal{A}=\{|F\rangle,|C_1\rangle,|C_2\rangle,|C_3\rangle\}=\{|K\uparrow\rangle,|K\downarrow\rangle,|K'\uparrow\rangle,|K'\downarrow\rangle\}$ as shown in Fig. \ref{fig:LLmodes}. The spinors $|C_i\rangle$ define the empty sub-LLs of the basis $\mathcal{A}$. In this basis, we are able to define the six broken generators
\begin{equation}
\begin{aligned}[c]
\Gamma^1_x&=\frac{1}{2}\sigma_xP_{+n_z}\\
\Gamma^2_x&=\frac{1}{2}\tau_xP_{+s_z}\\
\Gamma^3_x&=\frac{1}{4}(\sigma_x\tau_x-\sigma_y\tau_y)
\end{aligned}\quad
\begin{aligned}[c]
\Gamma^1_y&=\frac{1}{2}\sigma_yP_{+n_z}\\
\Gamma^2_y&=\frac{1}{2}\tau_yP_{+s_z}\\
\Gamma^3_y&=\frac{1}{4}(\sigma_x\tau_y+\sigma_y\tau_x),
\end{aligned}
\label{eq:broken_generators}
\end{equation}
where $P_{+s_z}=\frac{1}{2}(1+\sigma_z)$ and $P_{+n_z}=\frac{1}{2}(1+\tau_z)$ are the projectors over the spin up and valley $K$, respectively. Here, the matrices $\bm{\sigma}$ and $\bm{\tau}$ are the usual Pauli matrices acting in the spin and pseudo-spin spaces, respectively. Explicitely, the $\Gamma_x$ operators are
\begin{align}
\Gamma^1_x&=\begin{pmatrix}
0 &1&0&0\\1&0&0&0\\0&0&0&0\\0&0&0&0
\end{pmatrix} \quad
\Gamma^2_x=\begin{pmatrix}
0 &0&1&0\\0&0&0&0\\1&0&0&0\\0&0&0&0
\end{pmatrix} \nonumber \\
\Gamma^3_x&=\begin{pmatrix}
0 &0&0&1\\0&0&0&0\\0&0&0&0\\1&0&0&0
\end{pmatrix}.
\end{align}
The matrices $\Gamma^1_{x,y}$ mix $|F\rangle$ and $|C_1\rangle$, the matrices $\Gamma^2_{x,y}$ mix $|F\rangle$ and $|C_2\rangle$ while the matrices $\Gamma^3_{x,y}$ mix $|F\rangle$ and $|C_3\rangle$. We have thus three sets of canonically conjugate matrices such that for each mode $a$
\begin{align}
	[\Gamma_\mu^a,\Gamma_\nu^a]&=i\varepsilon_{\mu\nu\lambda}\Gamma_\lambda^a \\
	\{\Gamma_\mu^a,\Gamma_\nu^a\}&=\frac{i}{2}\delta_{\mu\nu},
\end{align}
where $\mu,\nu,\lambda\in\{x,y,z\}$, $a\in\{1,2,3\}$, $\varepsilon_{\mu\nu\lambda}$ is the three-dimensional Levi-Civita tensor, $\delta_{\mu\nu}$ is the identity matrix and we have introduced the additional matrices
\begin{align}
\Gamma^1_z=\frac{1}{2}\sigma_zP_{+n_z},\quad
\Gamma^2_z=\frac{1}{2}\tau_zP_{+s_z}, \quad
\Gamma^3_z=\frac{1}{4}(\sigma_z+\tau_z),
\end{align}
to complete the algebra. To study the spin waves for another phase, we simply rotate the spinors and the generators by a SU(4) unitary transformation $U$
\begin{subequations}
	\begin{align}
	&|\tilde{F}\rangle=U|F\rangle, \\
	&|\tilde{C}_i\rangle=U|C_i\rangle, \\
	&\tilde{\Gamma}_\mu^a=U\Gamma_\mu^aU^\dagger.
	\end{align}
	\label{eq:change_basis}
\end{subequations}

An important object that characterizes the spin waves in a (anti-)ferromagnet is the matrix of the commutators of the broken generators over the ground state
\begin{align}
M_{\mu\nu}^{ab}=\langle F|[\Gamma_{\mu}^a,\Gamma_{\nu}^b]|F\rangle,
\end{align}
with $\mu,\nu\in\{x,y\}$. We find that it is independent of the basis and defines the number and dispersion of the Goldstone modes associated with the number of broken symmetry\cite{Nielsen1976,Watanabe2011,Hidaka2013} (in the absence of explicit symmetry breaking terms). We find that this matrix has the expression for any phase
\begin{align}
\langle F|[\Gamma_{\mu}^a,\Gamma_{\nu}^b]|F\rangle=\frac{i}{2}\varepsilon_{\mu\nu}\delta_{ab}.
\label{eq:commutators}
\end{align}
where $\varepsilon_{\mu\nu}$ is the two-dimensional Levi-Civita tensor for $\mu,\nu\in\{x,y\}$. According to the general theory of Refs. [\onlinecite{Watanabe2011}] and [\onlinecite{Hidaka2013}], the number of quadratic spin waves is equal to $\text{Rank}[M]/2=3$ while no linearly dispersing modes are found which is in agreement with Refs. [\onlinecite{Arovas1999}] and [\onlinecite{Yang2006}], where the number of Goldstone modes is shown to be half the number of the broken symmetries because half of the fields are conjugate to the other half. We thus expect three quadratically dispersing modes in the absence of symmetry breaking terms. However, we show below that in some cases, the Goldstone modes become linear at small wavevectors due to spin-valley anisotropic terms that explicitely break the residual symmetry to yet lower ones.

\subsection{Lagrangian}

The effective low-energy Lagrangian is obtained analogously to Ref. [\onlinecite{Moon1995}] by constructing a coherent state 
\begin{align}
	|\psi[\pi]\rangle=e^{i\sum_{\mathbf{r}_i}O(\mathbf{r}_i,t)}|\psi_0\rangle,
\end{align}
where $|\psi_0\rangle$ is the second quantized QHFM ground state (\ref{eq:ground_state}) and
\begin{align}
	O(\mathbf{r}_i,t)=\pi_\mu^a(\mathbf{r}_i,t)\Gamma_\mu^a(\mathbf{r}_i),
\end{align}
where $\pi_\mu^a(\mathbf{r}_i,t)$ are six real fields associated with the broken generators $\Gamma_\mu^a(\mathbf{r}_i)$ acting at the Landau site $\mathbf{r}_i$ and we have assumed summation over repeated indices. They correspond to generalized local spin-valley rotations and thus describe the quantum state $|\psi[\pi]\rangle$ with spin-valley textures.

The total Lagrangian $\mathcal{L}$ is the sum of the kinetic term $\mathcal{L}_K$, the Coulomb term $\mathcal{L}_C$ and the symmetry breaking $\mathcal{L}_{SB}$ terms 
\begin{align}
\mathcal{L}&=\mathcal{L}_K-\mathcal{L}_C-\mathcal{L}_{SB}, \\
\mathcal{L}_K&=\langle\psi[\pi]|i\partial_t|\psi[\pi]\rangle, \\
\mathcal{L}_C&=\langle\psi[\pi]|H_C|\psi[\pi]\rangle, \\
\mathcal{L}_{SB}&=\langle\psi[\pi]|H_A|\psi[\pi]\rangle-\langle\psi_0|H_A|\psi_0\rangle.
\end{align}
In order to derive the effective non-linear sigma model at low-energy, we follow closely Refs. [\onlinecite{Arovas1999}], [\onlinecite{Yang2006}] and [\onlinecite{Kharitonov2012}]. 

\subsubsection{Kinetic term}

In the continuum limit, the kinetic term can be expressed as
\begin{align}
	\mathcal{L}_K=\rho_0\int d^2rZ^\dagger(\mathbf{r},t)i\partial_tZ(\mathbf{r},t),
\end{align}
in terms of the spinor field
\begin{align}
Z(\mathbf{r},t)=e^{iO(\mathbf{r},t)}|F\rangle,
\end{align}
where $|F\rangle$ is the ground state spinor corresponding to Eq. (\ref{eq:ground_state}). Expanding $O(\mathbf{r},t)$ up to second order in the $\pi$ fields, with the help of Eq. (\ref{eq:commutators}), we obtain
\begin{align}
\mathcal{L}_K&=\frac{\rho_0}{2}\int d^2r\varepsilon_{\mu\nu}\pi_\mu^a\partial_t\pi_\nu^a \label{eq:kinetic_L} \\
&=\frac{\rho_0}{2}\int d^2r\bm{\mathcal{A}}^a[\bm{\pi}]\cdot\partial_t\bm{\pi}^a,
\end{align}
where $\rho_0=(2\pi l_B^2)^{-1}$ is the electron density, and $\bm{\mathcal{A}}^a[\bm{\pi}]=(-\pi_y^a, \pi_x^a,0)$ is the Berry connection associated with the mode $a$.

\subsubsection{Gradient term}

To lowest order in the spatial derivatives, the energy associated with the Coulomb Hamiltonian gives rises to a gradient term\cite{Arovas1999,Yang2006,Kharitonov2012}
\begin{align}
\mathcal{L}_\text{C}&=\rho_s\int d^2r\text{Tr}\left[\bm{\nabla}P\bm{\nabla}P\right]\\
&=2\rho_s\int d^2r\partial_jZ^\dagger(1-ZZ^\dagger)\partial_jZ,
\end{align}
where
\begin{align}
	P(\mathbf{r},t)=ZZ^\dagger
\end{align}
is the (space-time dependent) order parameter of the ferromagnet and
\begin{align}
	\rho_s=\frac{1}{16\sqrt{2\pi}}\frac{e^2}{\varepsilon l_B}
\end{align}
is the spin stiffness. This gradient term corresponds to the cost in exchange energy associated with the misalignment of neighboring spins. 

The matrix $P$ is a projector\cite{Kharitonov2012} that obeys $P^2=P$, $P^\dagger=P$ and $\text{Tr}[P]=1$. Up to second order in the $\pi$-fields, the gradient term is given by
\begin{align}
\mathcal{L}_C=\frac{\rho_s}{2}\int d^2r(\bm{\nabla}\pi_\mu^a)^2,
\label{eq:Coulomb_L}
\end{align}
where we have used the property that $\langle F|\Gamma_\mu^a\Gamma_\nu^b|F\rangle=\frac{1}{4}\delta_{ab}(\delta_{\mu\nu}+i\varepsilon_{\mu\nu})$. We recover thus the usual non-linear sigma model term extended to the six fields in the $CP^3$ space.

\subsubsection{Anisotropic terms}

Finally, the symmetry breaking terms correspond to the anisotropic energy $E_A[Z]$ of the slowly varying field $Z$ minus the anisotropic energy of the ground state such that we consider only the \textit{excess} energy corresponding to the spin wave
\begin{align}
	\mathcal{L}_A=E_A[Z]-E_A[F],
\end{align}
where $E_A[F]$ is given by Eq. (\ref{eq:aniso_GS}) and
\begin{align}
	E_A[Z]=\rho_0\int d^2r\Big\{\sum_iu_i M_{P_i}^2[Z]-\Delta_ZM_{S_z}[Z]\Big\},
\end{align}
with $i\in\{x,y,z\}$, $u_x=u_y=u_\perp$, and
\begin{align}
\mathbf{M_{P}}[Z]&=\langle Z|\bm{\tau}|Z\rangle  \\
\mathbf{M_{S}}[Z]&=\langle Z|\bm{\sigma}|Z\rangle
\end{align}
are the spin and pseudo-spin magnetizations analogous to (\ref{eq:spin_mag_GS}) generalized to the field $Z$. We can express the anisotropic Lagrangian in a more compact way
\begin{align}
	\mathcal{L}_A=\rho_0\int d^2r\sum_i u_it_i-\Delta_Zs_z,
\end{align}
with
\begin{align}
	t_i&=\langle Z|\tau_i|Z\rangle^2-\langle F|\tau_i|F\rangle^2 \\
	s_z&=\langle Z|\sigma_z|Z\rangle-\langle F|\sigma_z|F\rangle.
\end{align}
We now expand the pseudo-spin magnetization up to second order in the $\pi$-fields
\begin{align}
	\langle Z| \tau_i |Z\rangle&=\langle F|e^{-iO}\tau_i e^{iO}|F\rangle \nonumber \\
	&=\langle F| \tau_i |F\rangle-i\pi_\mu^a\langle F|[\Gamma_\mu^a,\tau_i]|F\rangle\nonumber \\
	&-\frac{1}{2}\pi_\mu^a\langle F|[\Gamma_\mu^a,[\Gamma_\nu^b,\tau_i]]|F\rangle\pi_\nu^b,
	\label{eq:pseudo-spin_mag}
\end{align}
and we have a similar expression for the spin magnetization. Upon squaring, the pseudo-spin anisotropy has a linear and a quadratic term in the $\pi$-fields 
\begin{align}
t_i=R_{\mu}^{0a}\pi_\mu^a+\pi_\mu^aR_{i,\mu\nu}^{ab}\pi_\nu^b,
\end{align}
with
\begin{align}
R_{i\mu}^{0a}=&-2i\langle F|\tau_i|F\rangle\langle F|[\Gamma_\mu^a,\tau_i]|F\rangle\\
R_{i,\mu\nu}^{ab}=&-\langle F|[\Gamma^a_\mu,\tau_i]|F\rangle\langle F|[\Gamma^b_\nu,\tau_i]|F\rangle \nonumber \\
&-\langle F| \tau_i |F\rangle\langle F|[\Gamma_\mu^a,[\Gamma^b_\nu,\tau_i]]|F\rangle.
\label{eq:SB_pseudo}
\end{align}
The Zeeman term is linear in the spin magnetization such that we have
\begin{align}
s_z=R_{Z\mu}^{0a}\pi_\mu^a+\pi_\mu^aR_{Z,\mu\nu}^{ab}\pi_\nu^b,
\end{align}
where
\begin{align}
R_{Z\mu}^{0a}&=-i\langle F|[\Gamma_\mu^a,\sigma_z]|F\rangle \\
R_{Z,\mu\nu}^{ab}&=-\frac{1}{2}\langle F|[\Gamma_\mu^a,[\Gamma^b_\nu,\sigma_z]]|F\rangle
\label{eq:SB_spin}
\end{align}
For every state $|F\rangle$, the linear terms cancel each other
\begin{align}
\sum_iu_iR_{i\mu}^{0a}-\Delta_ZR_{Z\mu}^{0a}=0
\end{align}
for all $\mu$ and $a$. The anisotropic Lagrangian can thus be written as
\begin{align}
	\mathcal{L}_A=\int d^2r\bm{\pi}^T\mathcal{R}\bm{\pi}
\end{align}
where $\bm{\pi}=(\pi_\mu^a)$ is the six-component vector made of the $\pi$-fields and
\begin{align}
	\mathcal{R}_{\mu\nu}^{ab}=\sum_i u_iR_{i\mu\nu}^{ab}-\Delta_ZR_{Z\mu\nu}^{ab}
	\label{eq:symmetry_breaking}
\end{align} 
is a $6\times6$ matrix in the basis $\{\mu,a\}$ that we call the anisotropy matrix.

We now consider the effective action $\mathcal{S}=\int dt\mathcal{L}$ and Fourier transform the kinetic and gradient Lagrangians (\ref{eq:kinetic_L}) and (\ref{eq:Coulomb_L}) in space and time 
\begin{align}
	\mathcal{S}=\int d\omega d^2k\bm{\pi}^T(\mathbf{k},\omega)\mathcal{M}\bm{\pi}(-\mathbf{k},-\omega),
\end{align}
with
\begin{align}
\mathcal{M}_{\mu\nu}^{ab}=\left(\frac{\rho_0}{2}i\omega\varepsilon_{\mu\nu}-\frac{\rho_s}{2}\mathbf{k}^2\delta_{\mu\nu}\right)\delta_{ab}-\rho_0\mathcal{R}_{\mu\nu}^{ab}.
\label{eq:total_matrix}
\end{align}
The dispersion relations of the collective mode are obtained by minimizing the action, $\delta\mathcal{S}/\delta \mathbf{\pi}(\mathbf{k},\omega)=0$, which gives the equation
\begin{align}
	\mathcal{M}(\mathbf{k},\omega)\bm{\pi}(\mathbf{k},\omega)=0 .
\end{align}
Because the matrix $\mathcal{M}(\mathbf{k},\omega)$ is hermitian, the frequencies always come in pairs $\pm\omega(\mathbf{k})$. However, we only consider the three positive eigenfrequencies $\omega_\alpha$($\mathbf{k}$), which correspond to the physically relevant modes, and discard the negative-energy solutions. The corresponding fields $\bm{\pi}$ are obtained by finding the null space of $\mathcal{M}$. The resulting spinor is thus given by
\begin{align}
	|Z_\alpha\rangle=\left(\mathbb{1}+i\pi_{\mu,\alpha}^a\Gamma_\mu^a-\frac{1}{2}\pi_{\mu,\alpha}^a\pi_{\nu,\alpha}^b\Gamma_\mu^a\Gamma_\nu^b\right)|F\rangle,
\end{align}
where $\pi_{\mu\alpha}^a$ is the eigenstate corresponding to the frequency $\omega_\alpha$. When the matrix is block-diagonal $\mathcal{M}_{\mu\nu}^{ab}\propto\delta_{ab}$, the different modes are decoupled and the eigenstate labels are identical to the mode label $\alpha=a$. This is the case for the CDW and KD phases.

\subsection{Change of ground state}

The general analysis of the previous sections has been performed by considering the ground state spinor $|F\rangle=|\mathbf{n}_z\rangle|\mathbf{s}_z\rangle$. To consider a different ground state, we perform the unitary rotation given by Eqs. (\ref{eq:change_basis}). The spinor $Z$ is thus transformed as
\begin{align}
	\tilde{Z}=UZ=e^{i\tilde{\pi}_\mu^a\tilde{\Gamma}_\mu^a}|\tilde{F}\rangle,
\end{align}
where we have introduced the the fields $\tilde{\pi}_\mu^a$ which correspond now to the modes $a$ associated with the broken generators $\tilde{\Gamma}_\mu^a$. However, for simplicity, we will keep the notation $\pi_\mu^a$ in every basis and assume that the $\pi$-fields correspond to the modes in the corresponding basis.

The kinetic and gradient terms are independent of the basis because the SU(4) transformation matrix $U$ is global $\mathcal{L}_K[\tilde{Z}]=\mathcal{L}_K[Z]$ and $\mathcal{L}_C[\tilde{Z}]=\mathcal{L}_C[Z]$. However, the symmetry breaking terms are basis dependent. The spin and pseudo-spin magnetization read
\begin{align}
	\langle\tilde{Z}|\bm{\tau}|\tilde{Z}\rangle&=\langle Z|\mathbf{P}|Z\rangle \\
	\langle\tilde{Z}|\sigma_z|\tilde{Z}\rangle&=\langle Z|S_z|Z\rangle,
\end{align}
such that instead of computing the commutators in Eq. (\ref{eq:pseudo-spin_mag}) using the transformed matrices $\tilde{\Gamma}_\mu^a$, we simply replace the matrices $\bm{\tau}$ and $\sigma_z$ by 
\begin{align}
	\mathbf{P}&=U^\dagger\bm{\tau}U \\
	S_z&=U^\dagger\sigma_z U,
\end{align}
such that the pseudo-spin magnetization reads
	\begin{align}
\langle \tilde{Z}| \tau_i |\tilde{Z}\rangle&=\langle F| P_i |F\rangle-i\pi_\mu^a\langle F|[\Gamma_\mu^a,P_i]|F\rangle\nonumber \\
&-\frac{1}{2}\pi_\mu^a\langle F|[\Gamma_\mu^a,[\Gamma_\nu^b,P_i]]|F\rangle\pi_\nu^b,
\end{align}
where $|F\rangle=|\mathbf{n}_z\rangle|\mathbf{s}_z\rangle$ and the matrices $\Gamma_\mu^a$ are given by Eqs. (\ref{eq:broken_generators}). We have a similar expression for the spin magnetization in the transformed basis. Thus instead of computing the transformed matrices and spinors in the new basis, we simply express the matrices $\bm{\tau}$ and $\sigma_z$ in the basis $\tilde{\mathcal{A}}$. Thus the anisotropic Lagrangian reads
\begin{align}
\mathcal{L}_A[\tilde{Z}]=\int d^2r\bm{\pi}\tilde{\mathcal{R}}\bm{\pi},
\end{align}
where
\begin{align}
\tilde{\mathcal{R}}_{\mu\nu}^{ab}=\sum_i u_i\tilde{R}_{i\mu\nu}^{ab}-\Delta_Z\tilde{R}_{Z\mu\nu}^{ab},
\end{align} 
and the matrices $\tilde{R}_{i\mu\nu}^{ab}$ and $\tilde{R}_{Z\mu\nu}^{ab}$ are obtained from Eqs. (\ref{eq:SB_pseudo}) and (\ref{eq:SB_spin}) by the replacements $\tau_i\rightarrow P_i$ and $\sigma_z\rightarrow S_z$.

\section{Dispersion relations}

\label{sec:Disp}

Using the formalism developped in the previous section, we now diagonalize the matrix (\ref{eq:total_matrix}) to find the dispersion relations of the three different modes and their associated gaps. We only consider the four phases of Sec. \ref{sec:phase_diag_without} without the valley Zeeman term since they are not substantially modified upon its introduction.

\subsection{Charge density wave phase}

In the charge density wave, the ground state spinor and the empty sub-LL $|C_a\rangle$ defining the three mode $a$ have the expression 
\begin{subequations}
	\begin{align}
	&|F\rangle=|\mathbf{n}_z\rangle|\mathbf{s}_z\rangle=(1,0,0,0)^T \\
	&|C_1\rangle=|\mathbf{n}_z\rangle|-\mathbf{s}_z\rangle=(0,1,0,0)^T \\
	&|C_2\rangle=|-\mathbf{n}_z\rangle|\mathbf{s}_z\rangle=(0,0,1,0)^T \\
	&|C_3\rangle=|-\mathbf{n}_z\rangle|-\mathbf{s}_z\rangle=(0,0,0,1)^T
	\end{align}
	\label{eq:spinorsCDW}
\end{subequations}
in the basis $\{|K\uparrow\rangle,|K\downarrow\rangle,|K'\uparrow\rangle,|K'\downarrow\rangle\}$. We have chosen here a ground state polarized in valley $K$, but one can also choose a polarization in valley $K'$ by the replacement $\mathbf{n}_z\rightarrow-\mathbf{n}_z$. The mode $a=1$ which mixes $|F\rangle$ and $|C_1\rangle$ corresponds to a pure spin wave such that the pseudo-spin remains unaffected. The mode $a=2$ mixes $|F\rangle$ and $|C_2\rangle$ and corresponds to a pseudo-spin wave where the spin remains unaffected. The mode $a=3$ corresponds to an entanglement wave in which inverses both the spin and pseudo-spin such that the spinor $Z$ is in a superposition of $|\mathbf{n}_z\rangle|\mathbf{s}_z\rangle$ and $|-\mathbf{n}_z\rangle|-\mathbf{s}_z\rangle$.

\begin{figure}[t]
	\begin{center}
		\includegraphics[width=8cm]{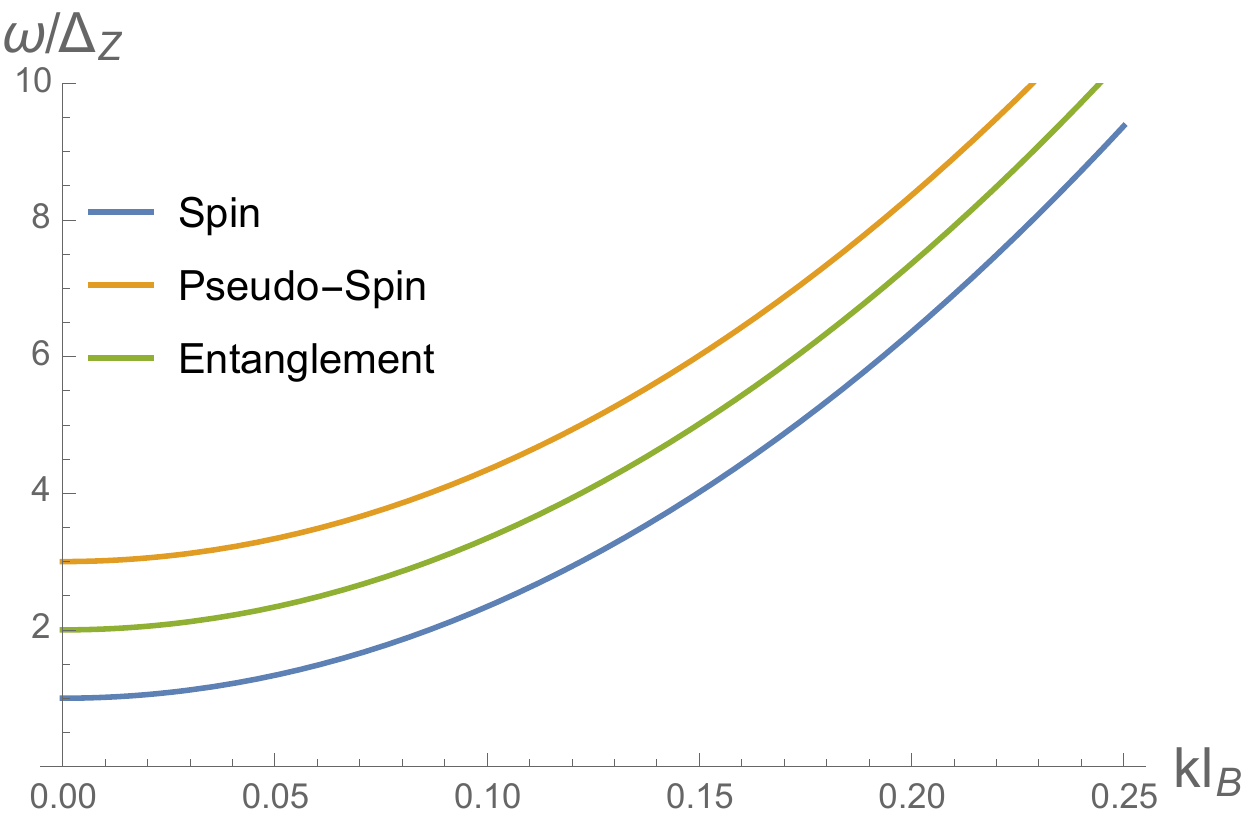} 
		\caption{Dispersion relation of the three modes in the KD phase for $u_z=-\Delta_Z$ and $u_\perp=2\Delta_Z$. The three modes are gapped and quadratically dispersing. }
		\label{fig:DispCDW}
	\end{center}
\end{figure}

\begin{figure}[t]
	\begin{center}
		\includegraphics[width=8cm]{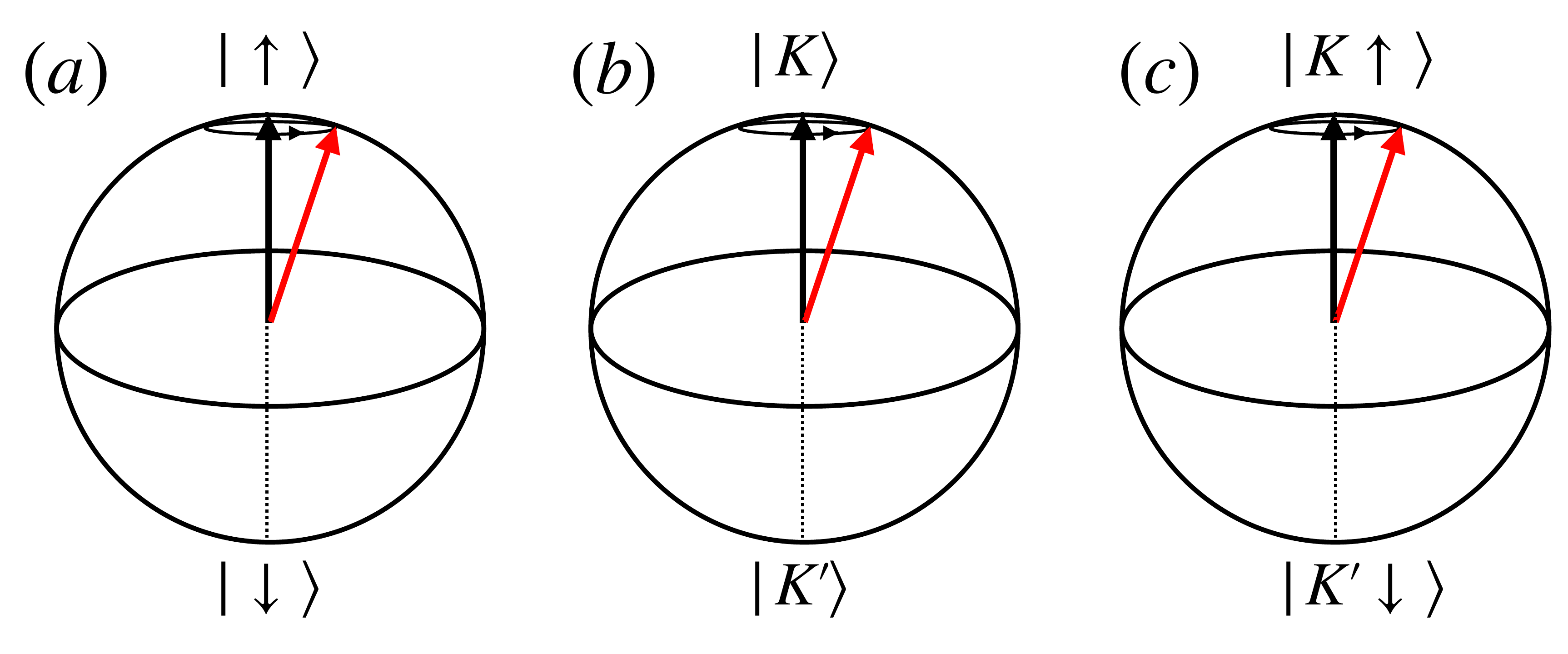} 
		\caption{Bloch spheres corresponding to each modes in the CDW phase. (a) Spin Bloch sphere of the pure spin mode 1. (b) Pseudo-spin Bloch sphere of the pseudo-spin mode 2. (c) Entanglement Bloch sphere corresponding to the entanglement mode 3. The black arrow indicates the ground state polarization, while the red arrow corresponds to the magnetization at a point in space in the presence of a spin wave. The red arrow rotates periodically around the ground state polarization according to Eq. (\ref{eq:space-time_spinors}).}
		\label{fig:BlochSpheresCDW}
	\end{center}
\end{figure}

The anisotropy matrix $\mathcal{R}$ is block diagonal $\mathcal{R}^{ab}_{\mu\nu}\propto\delta_{ab}$ such that the three modes are decoupled. We find the dispersion relations $\omega_a(\mathbf{k})$ corresponding to the three modes $a={1,2,3}$
\begin{align}
\omega_1(\mathbf{k})&=2\pi\rho_s(\mathbf{k}l_B)^2+\Delta_Z \\
\omega_2(\mathbf{k})&=2\pi\rho_s(\mathbf{k}l_B)^2+u_\perp-u_z \label{eq:pseudo-spinCDW}\\
\omega_3(\mathbf{k})&=2\pi\rho_s(\mathbf{k}l_B)^2+\Delta_Z-u_z.
\end{align}
The three modes have a quadratic dispersion and a mass term proportional to the anisotropic energy terms. The CDW region is defined by $u_\perp>u_z$ and $u_z<\Delta_Z$ such that the three modes have a positive gap in the region. The three eigenmodes have the same expression for each mode such that the spinor with wavevector $\mathbf{k}$ corresponding to mode $a$ reads
\begin{align}
	Z_{\mathbf{k}a}(\mathbf{r},t)=\left(1-\frac{\pi_0^2}{8}\right)|F\rangle+i\frac{\pi_0}{2}e^{i(\mathbf{k}\cdot\mathbf{r}-\omega_a t)}|C_a\rangle
	\label{eq:space-time_spinors}
\end{align}
where $\pi_0\ll 1$ is the magnitude of the wave which is a scalar. As shown in Fig. \ref{fig:BlochSpheresCDW}, The spin wave corresponds thus to a small weight on the spinor $|C_a\rangle$ with the phase oscillating at frequency $\omega_a$ and wavevector $\mathbf{k}$.

\begin{figure}[t]
	\begin{center}
		(a)\includegraphics[width=4.8cm]{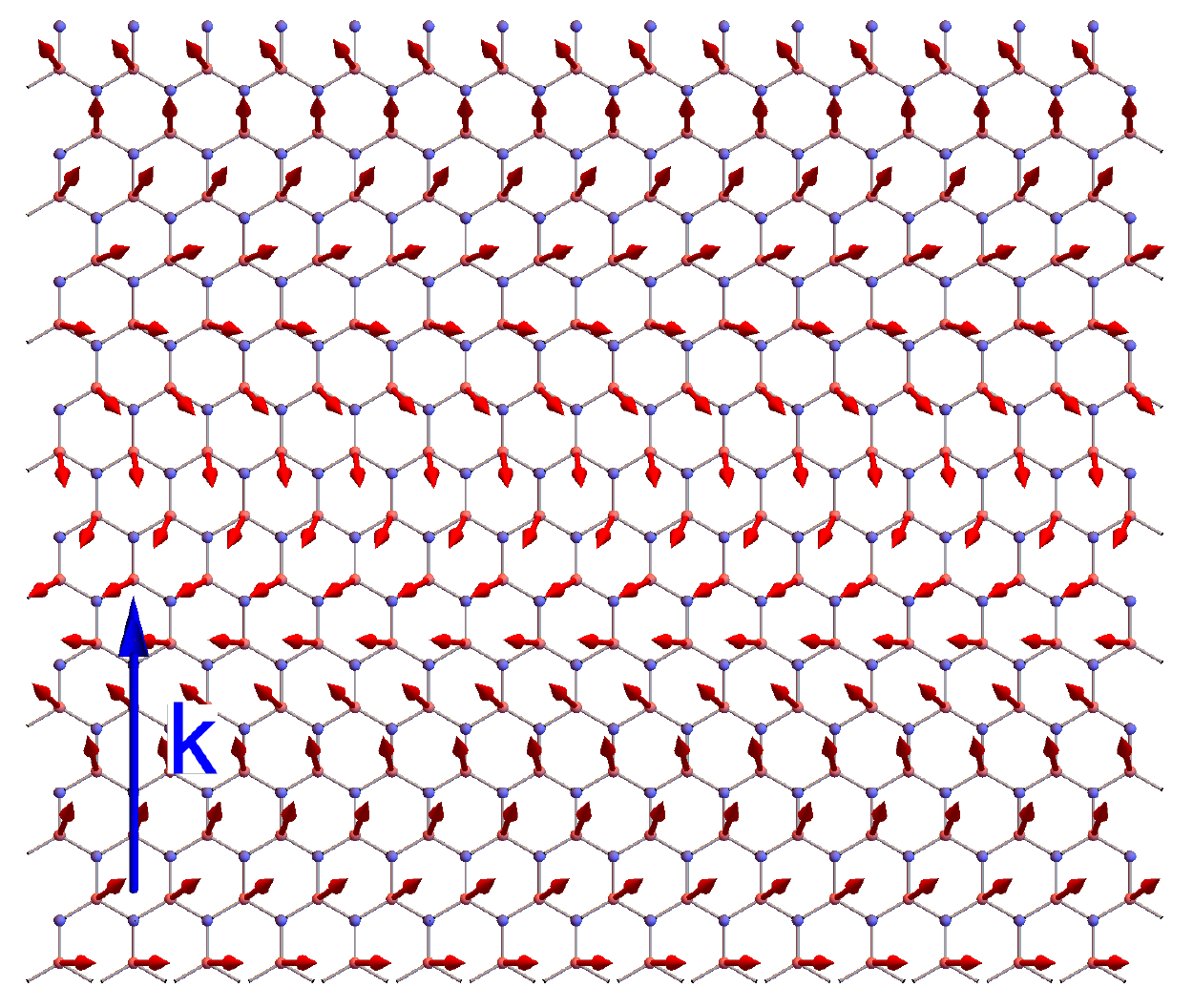} 
		(b)\includegraphics[width=3.6cm]{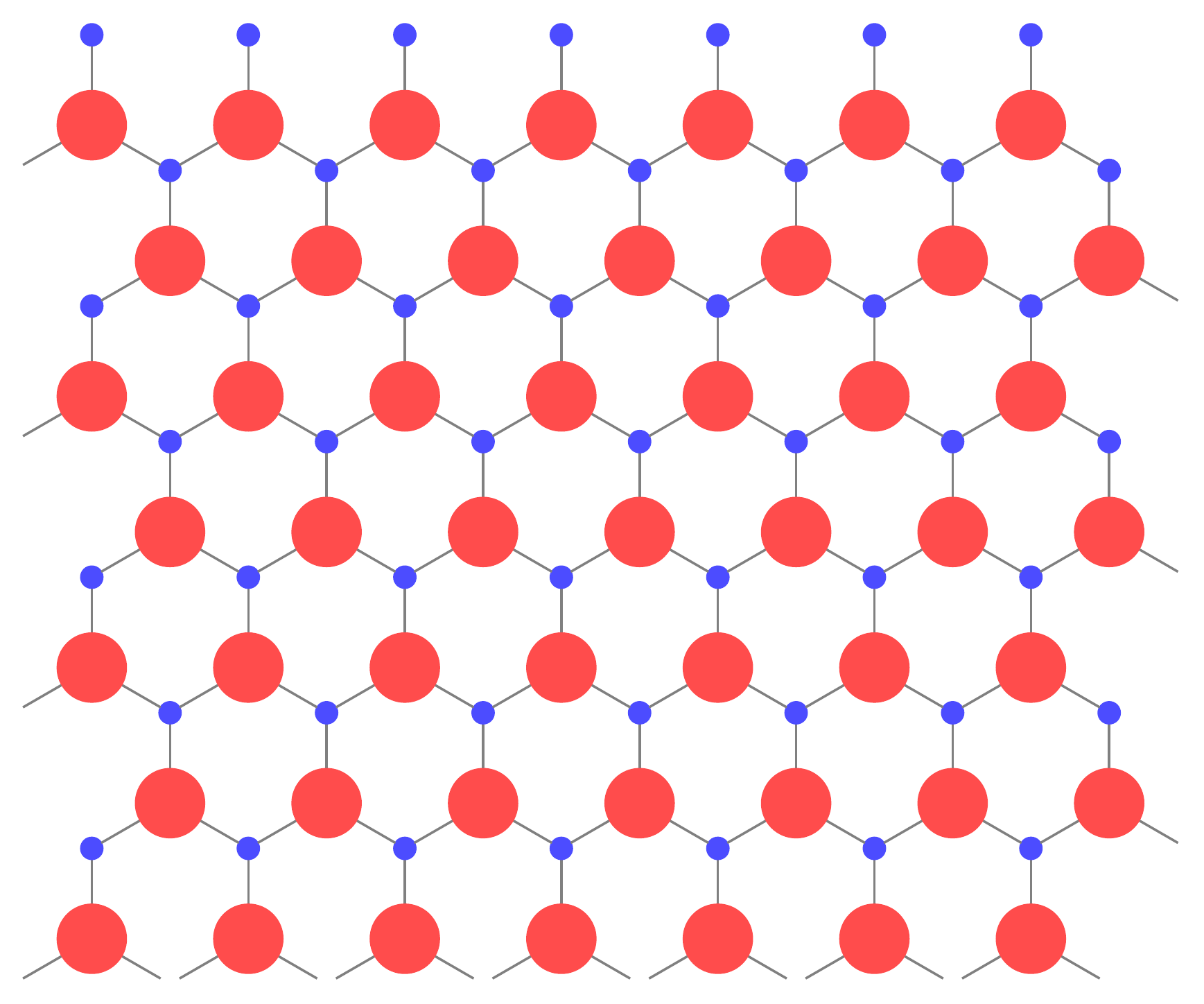}
		(c)\includegraphics[width=4.1cm]{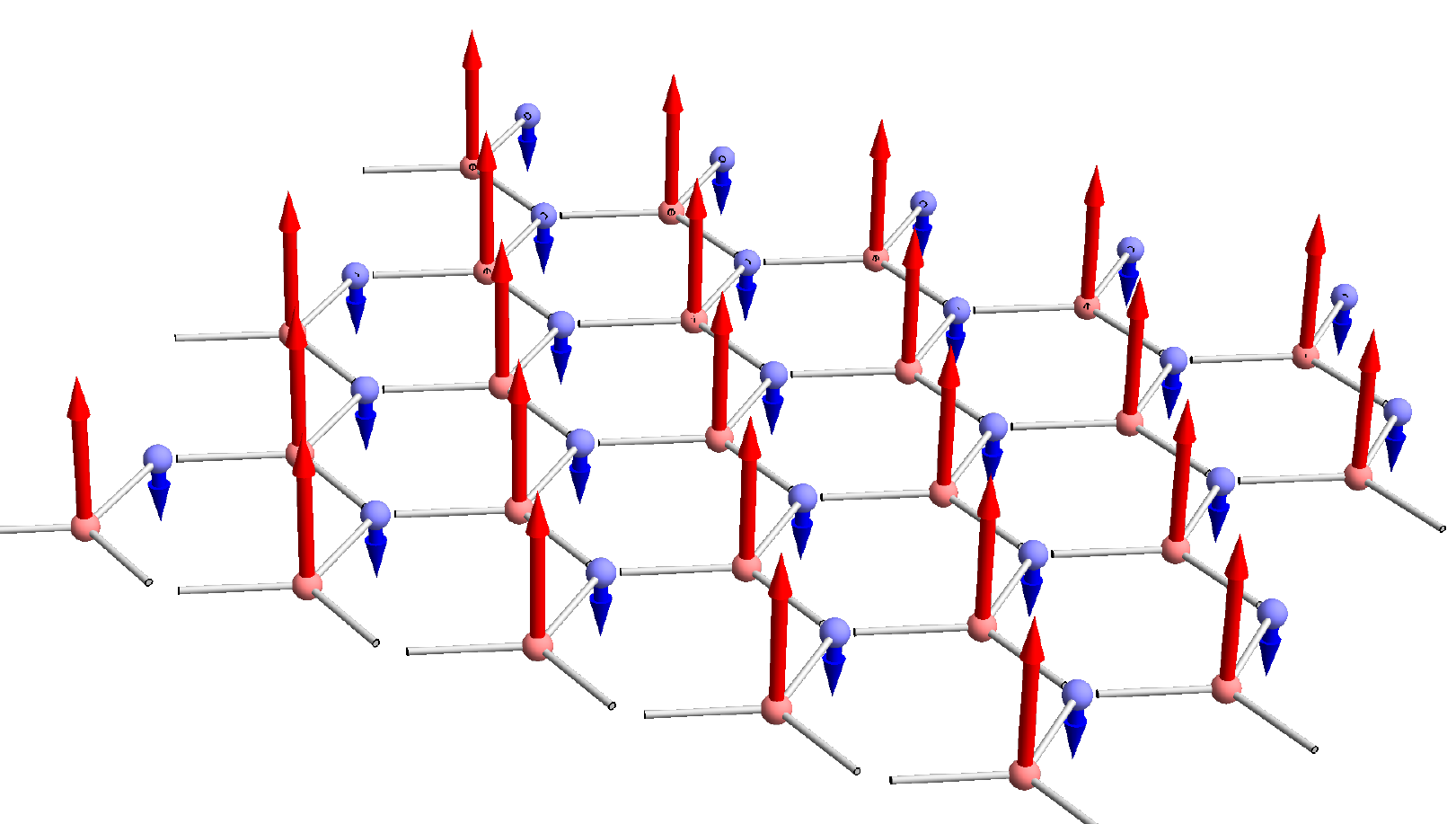}
		\caption{Three modes of the CDW phase. (a) "Snapshot" of the pure spin wave mode $a=1$ seen from the top with wavevector $\mathbf{k}$ along the axis $y$. We observe the precession of the spins around the $z$ axis of the spins in the A sublattice. (b) Sublattice polarization of the pseudo-spin wave mode $a=2$. We observe a small electronic density on sublattice B. The dynamic part of the field is encoded in the relative phase of the superposition between valley $K$ and $K'$. The spin magnetization is proportional to the sublattice density and points along $\mathbf{s}_z$. (c) Spin magnetization on the A and B sublattices of the entanglement mode $a=3$, there is a small spin magnetization on the B sublattice with opposite direction as on sublattice A.}
		\label{fig:modesCDW}
	\end{center}
\end{figure}

The first mode corresponds to a pure spin wave, its gap is unaffected by the anisotropic term and depends only on the Zeeman term. Because the pseudo-spin remains unaffected by the spin wave and is polarized in one valley, the spins live only on one sublattice (we choose sublattice A here for illustration) and spin the magnetization is
\begin{align}
	\mathbf{M_{S_A}}=\begin{pmatrix}
	\pi_0\cos(\mathbf{k}\cdot\mathbf{r}-\omega_1t) \\ \pi_0\sin(\mathbf{k}\cdot\mathbf{r}-\omega_1t)\\	1-\frac{1}{2}\pi_0^2
	\end{pmatrix} ,
	\quad \mathbf{M_{S_B}}=0.
	\label{eq:spin_mag}
\end{align}
The spin wave consist thus of the spins of sublattice A precessing around the axis $z$ at frequency $\omega_1$ as shown in Fig. \ref{fig:modesCDW}.(a).

The second mode corresponds to a pseudo-spin wave for which the gap depend only on the pseudo-spin anisotropic terms and not on the Zeeman term. In the CDW region, we have chosen for simplicity a polarization in the valley $K$ (a similar treatment can be done if the polarization is in valley $K'$) and thus the pseudo-spin points towards the north pole of the Bloch sphere. Because the pseudo-spin magnetization points along the $z$ direction, the anisotropic energy of the ground state depends only on $u_z$. The presence of a pseudo-spin wave introduces a pseudo-spin magnetization in the $xy$ plane of the Bloch sphere, such that the magnetization out-of-plane anisotropic energy $u_z$ is reduced, while there is a cost in in-plane anisotropic energy $u_\perp$, hence the gap is proportional to $(u_\perp-u_z)$. The pseudo-spin magnetization is given by
\begin{align}
\mathbf{M_{P}}=\begin{pmatrix}
\pi_0\cos(\mathbf{k}\cdot\mathbf{r}-\omega_2t) \\ 
\pi_0\sin(\mathbf{k}\cdot\mathbf{r}-\omega_2t)\\	
1-\frac{1}{2}\pi_0^2
\end{pmatrix} .
\end{align}
This expression for the pseudo-spin is analogous to the spin magnetization (\ref{eq:spin_mag}) of the pure spin wave. It is now the pseudo-spin that precesses around the $z$ axis, such that it corresponds to a superposition of the valley $K$ and $K'$ with a relative phase oscillating at frequency $\omega_2$. However, the electronic density imbalance of the sublattice, which corresponds to the $z$ component of the pseudo-spin magnetization ($M_{P_z}=\rho_A-\rho_B$) remains uniform
\begin{align}
	\rho_A=1-\frac{\pi_0^2}{4} \quad \rho_B=\frac{\pi_0^2}{4}
\end{align}
as shown in Fig. \ref{fig:modesCDW}.(b). We observe thus a small electronic density on the sublattice $B$. Because the spinors $|F\rangle$ and $|C_2\rangle$ both have spins pointing along the $z$ direction, the spin magnetization on sublattices A and B is simply proportional to the electronic density, $\mathbf{M_{S_A}}=\rho_A\mathbf{s}_z$ and $\mathbf{M_{S_B}}=\rho_B\mathbf{s}_z$. The total spin magnetization is thus $\mathbf{M_S}=\mathbf{s}_z$. 

The spinors of the third mode cannot be expressed as a tensor product of a spin and a valley spinors. Thereby, this mode is an entanglement mode which mixes the sub-LLs $|K\uparrow\rangle$ and $|K'\downarrow\rangle$. It corresponds to the electron being mainly polarized on sublatice A with spin up with a small polarization on sublatice B with spin down with the relative phase oscillating at frequency $\omega_3$. Analogously to the pseudo-spin wave, the pseudo-spin magnetization along the $z$ direction is reduced ($M_{P_z}=1-\pi_0^2/2$) such that there is a gain in anisotropic energy $u_z$. However, there is a cost in Zeeman energy, and the gap is proportional to $\Delta_Z-u_z$. The sublattice polarizarion is identical to the pseudo-spin wave but the spin magnetization is
\begin{align}
	\mathbf{M_{S_A}}=\left(1-\frac{\pi_0^2}{4}\right)\mathbf{s_z}\quad \mathbf{M_{S_B}}=-\frac{\pi_0^2}{4}\mathbf{s_z}
\end{align}
such that the total spin is reduced similarly to the spin wave. 

\begin{figure}[t]
	\begin{center}
		b)\includegraphics[width=3.6cm]{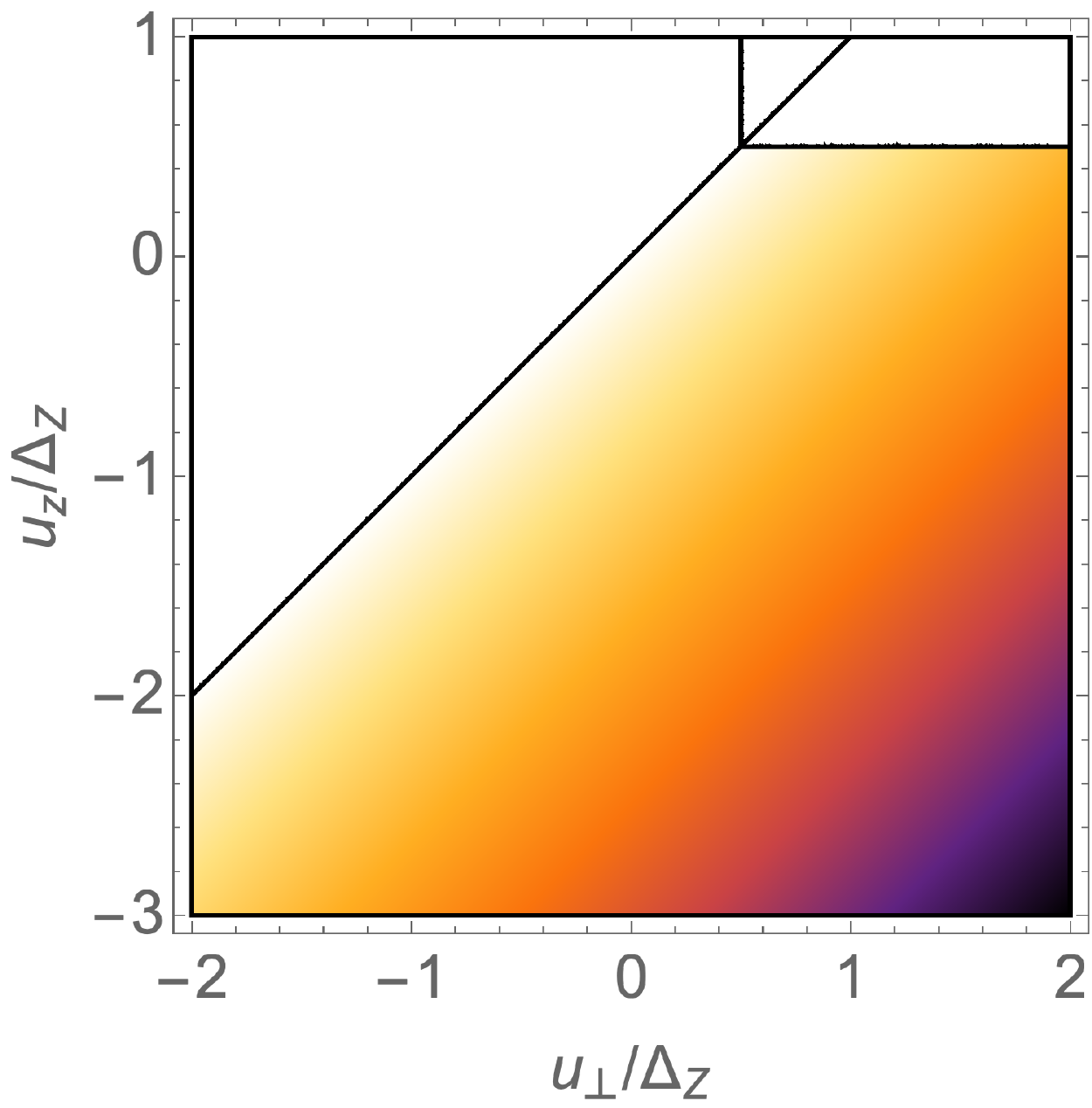}
		c)\includegraphics[width=4.2cm]{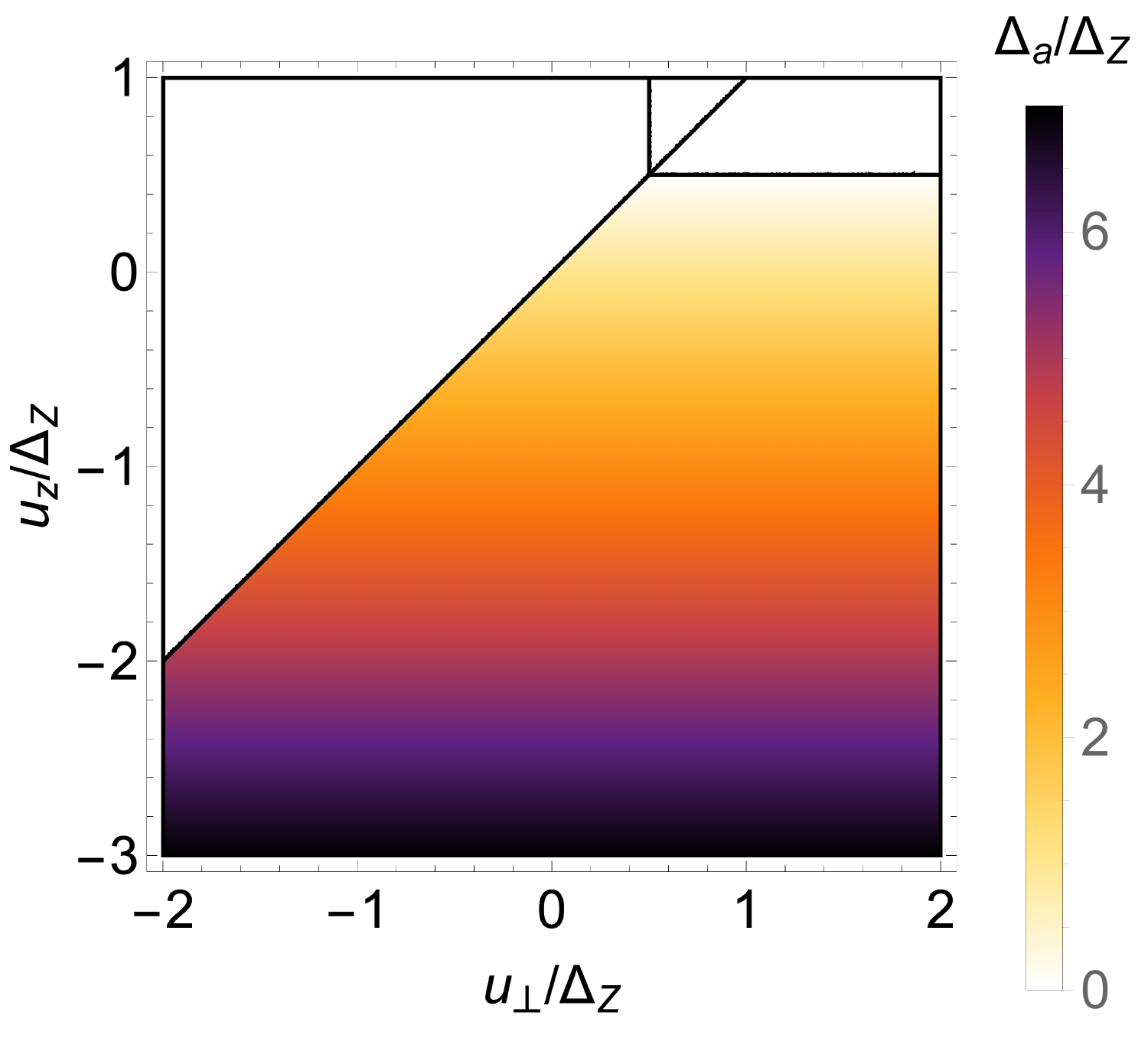}
		a)\includegraphics[width=5cm]{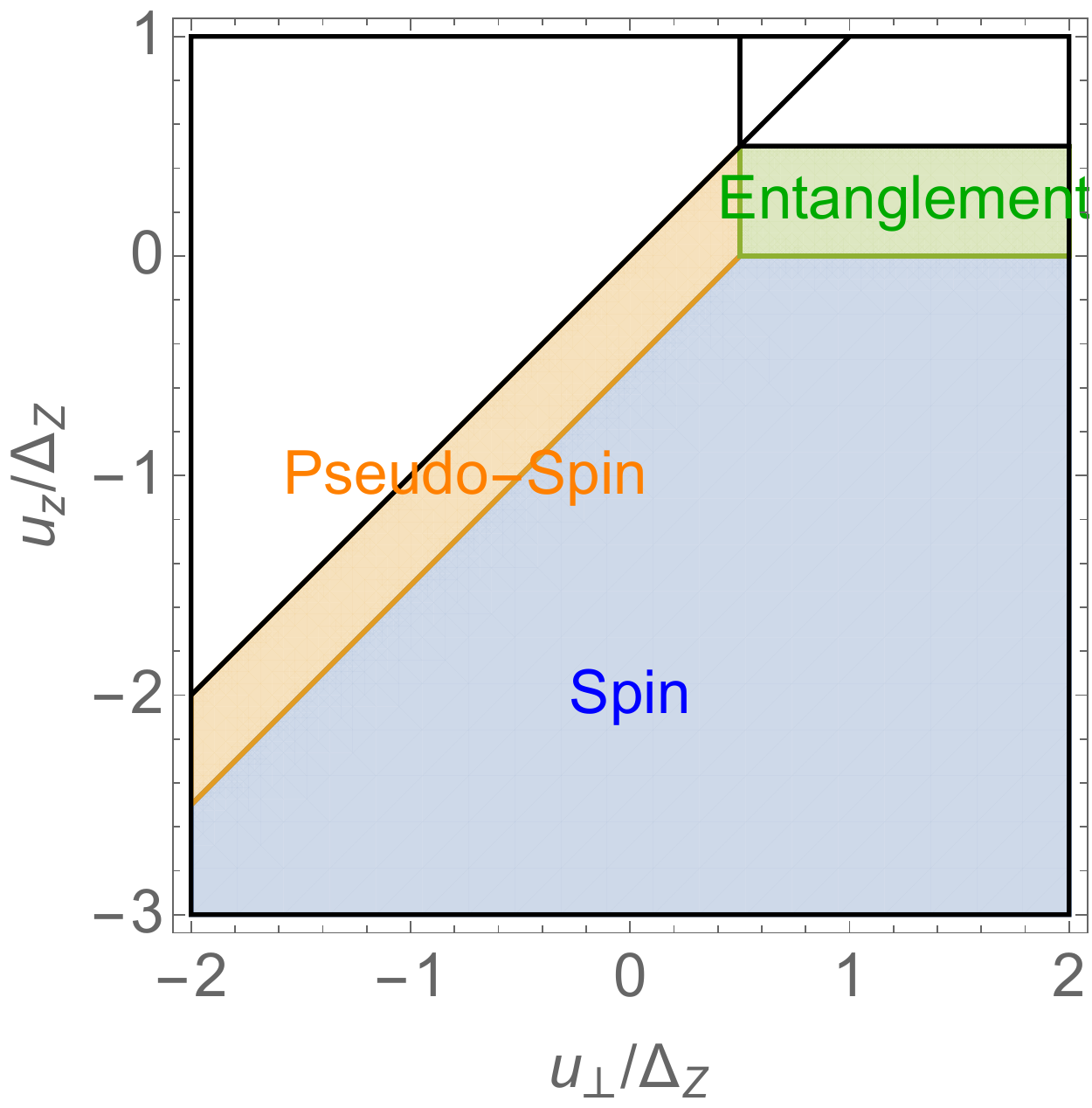} 
		\caption{Size of the gap of a) the pseudo-spin and b) the entanglement waves as a function of $u_\perp$ and $u_z$ in the CDW region. We observe that the pseudo-spin gap $\Delta_2$ vanishes at the boundary with the KD phase, and the entanglement gap $\Delta_3$ vanishes at the boundary with the AFI entangled phase. c) "Phase diagram" of the spin mode with the lowest gap. We can see that the pseudo-spin and entanglement modes have the lowest energy near the phase boundaries, whereas the spin mode dominates elsewhere.}
		\label{fig:GapsCDW}
	\end{center}
\end{figure}

Figs. \ref{fig:GapsCDW}.(a) and (b) show the size of the gaps $\Delta_a$ of the pseudo-spin and entanglement modes in units of $\Delta_Z$. We can see that the size of the gap decrease as we get closer to the boundaries and eventually vanish at the boundaries. 

The gap of the pseudo-spin wave vanishes at the boundary with the KD phase defined by $u_\perp=u_z$. At this line, as one can see from Eq. (\ref{eq:SB}), the SU(2) pseudo-spin symmetry is restored and there is thus no preferred orientation of the pseudo-spin. There is no cost in anisotropic energy for the creation of a pseudo-spin wave. The pseudo-spin wave becomes thus a true Goldstone mode where the spontaneously broken symmetry is the SU(2) pseudo-spin rotation symmetry. 

The gap of the entanglement wave vanishes at the boundary $u_z=\Delta_Z$ with the anti-ferrimagnetic phase which is an entangled phase. This comes from the fact that the spin and pseudo-spin magnetizations along $z$ of the wave are identical $M_{S_z}=M_{P_z}=(1-\pi_0^2/2)$ because we have a small imbalance over the state $|K'\downarrow\rangle$ with opposite spin \textit{and} pseudo-spin that of $|K\uparrow\rangle$. In addition, there is no spin and pseudo-spin magnetization in the $xy$ plane $M_{S_x,P_x}=M_{S_y,P_y}=0$. Thus, at the transition line $u_z=\Delta_Z$, up to second order in $\pi_0$, the anisotropic energy term
\begin{align}
	E_A[Z]=\frac{u_z}{2}M_{P_Z}^2-\Delta_ZM_{S_z}=\frac{u_z}{2}-\Delta_Z=E_A[F],
\end{align}
which is independent of the amplitude $\pi_0$. Thereby, for small amplitudes, the spin and pseudo-spin magnetizations cancel each other at the transition line. This symmetry between the spin and pseudo-spin magnetization will be explored further in Sec. \ref{sec:SWAFI}.

\subsection{Kekul\'e distortion phase}

In the KD phase, we apply the unitary transformation
\begin{align}
	U_{KD}=e^{i\frac{\pi}{4}\mathbf{n}\cdot\bm{\tau}},
	\label{eq:UKD}
\end{align}
with $\mathbf{n}=(\sin\varphi,-\cos\varphi,0)$ to the spinors (\ref{eq:spinorsCDW}) of the CDW phase such that we have the spinors in the KD phase
\begin{subequations}
	\begin{align}
	&|\tilde{F}\rangle=|\mathbf{n}_\perp\rangle|\mathbf{s}_z\rangle=\frac{1}{\sqrt{2}}(1,0,e^{i\varphi},0)^T \\
	&|\tilde{C}_1\rangle=|\mathbf{n}_\perp\rangle|-\mathbf{s}_z\rangle=\frac{1}{\sqrt{2}}(0,1,0,e^{i\varphi})^T \\
	&|\tilde{C}_2\rangle=|-\mathbf{n}_\perp\rangle|\mathbf{s}_z\rangle=\frac{1}{\sqrt{2}}(-e^{-i\varphi},0,1,0)^T \\
	&|\tilde{C}_3\rangle=|-\mathbf{n}_\perp\rangle|-\mathbf{s}_z\rangle=\frac{1}{\sqrt{2}}(0,-e^{-i\varphi},0,1)^T,
	\end{align}
	\label{eq:spinorsKD}
\end{subequations}
where we have a U(1) pseudo-spin symmetry in the $xy$ plane of the Bloch sphere. Similarly to the analysis for the CDW phase, the mode 1 is a pure spin wave where the pseudo-spin is unaffected, the mode 2 is a pseudo-spin wave, while the mode 3 is an entanglement mode.

\begin{figure}[t]
	\begin{center}
		\includegraphics[width=8cm]{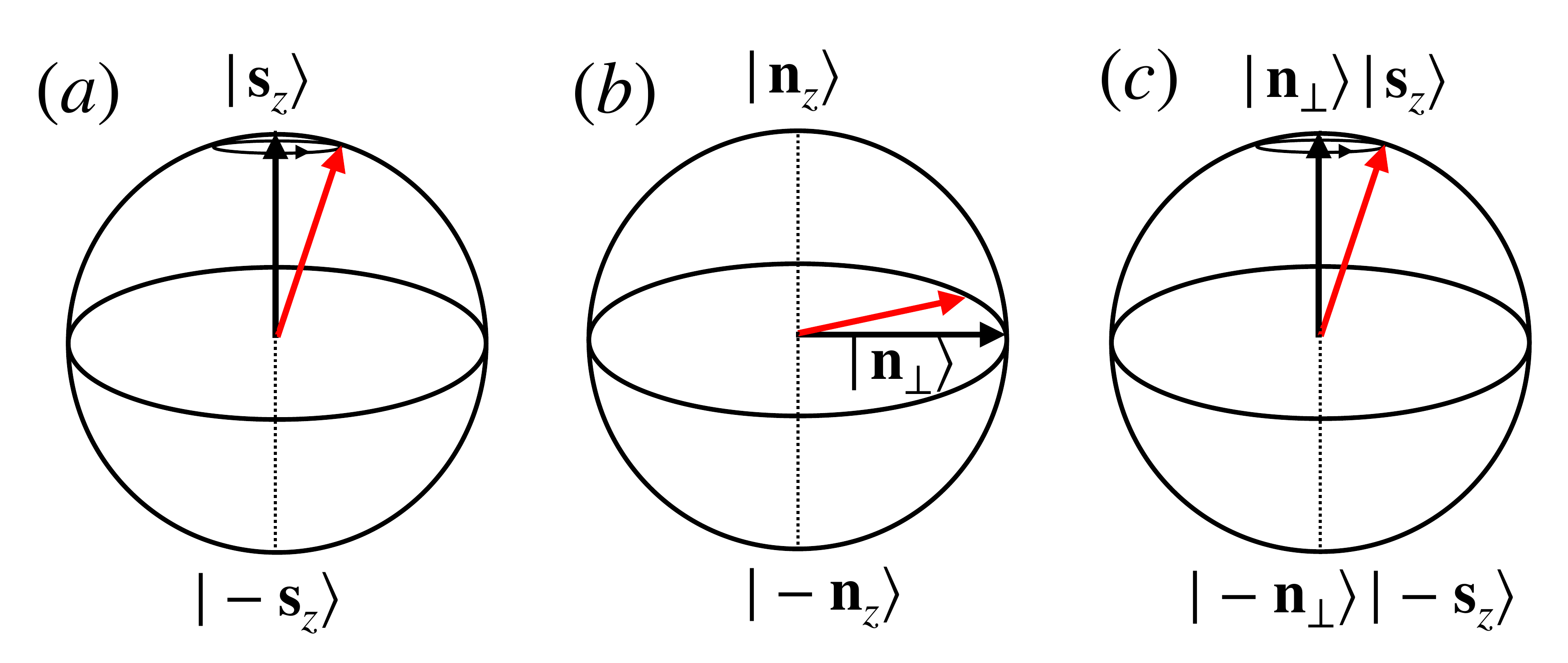} 
		\caption{Bloch spheres corresponding to each modes in the KD phase in the same way as Fig. \ref{fig:BlochSpheresCDW}. (a) Spin Bloch sphere of the spin mode 1. (b) Pseudo-spin Bloch sphere of the pseudo-spin mode 2. The ground state has a U(1) symmetry for rotations around the $z$ axis. (c) Entanglement Bloch sphere corresponding to the entanglement mode 3. For the spin and entanglement modes, the red arrow rotates periodically around the ground state polarization according to Eq. (\ref{eq:space-time_spinors}). At low-energy, the pseudo-spin mode is restricted to the equator of the pseudo-spin Bloch sphere, which costs no anisotropic energy, while at higher energy, it acquires an element along the $z$ direction.}
		\label{fig:BlochSpheresKD}
	\end{center}
\end{figure}

The anisotropy matrix $\mathcal{R}$ is again block diagonal $\mathcal{R}^{ab}_{\mu\nu}\propto\delta_{ab}$ such that the three modes are decoupled. We find the dispersion relations $\omega_a(\mathbf{k})$ corresponding to the three modes $a={1,2,3}$,
\begin{align}
\omega_1(\mathbf{k})&=2\pi\rho_s(\mathbf{k}l_B)^2+\Delta_Z \\
\omega_2(\mathbf{k})&=|\mathbf{k}|l_B\sqrt{2\pi\rho_s}\sqrt{2\pi\rho_s(\mathbf{k}l_B)^2+u_z-u_\perp} \\
\omega_3(\mathbf{k})&=2\pi\rho_s(\mathbf{k}l_B)^2+\Delta_Z-u_\perp.
\end{align} 
\begin{figure}[t]
	\begin{center}
		\includegraphics[width=8cm]{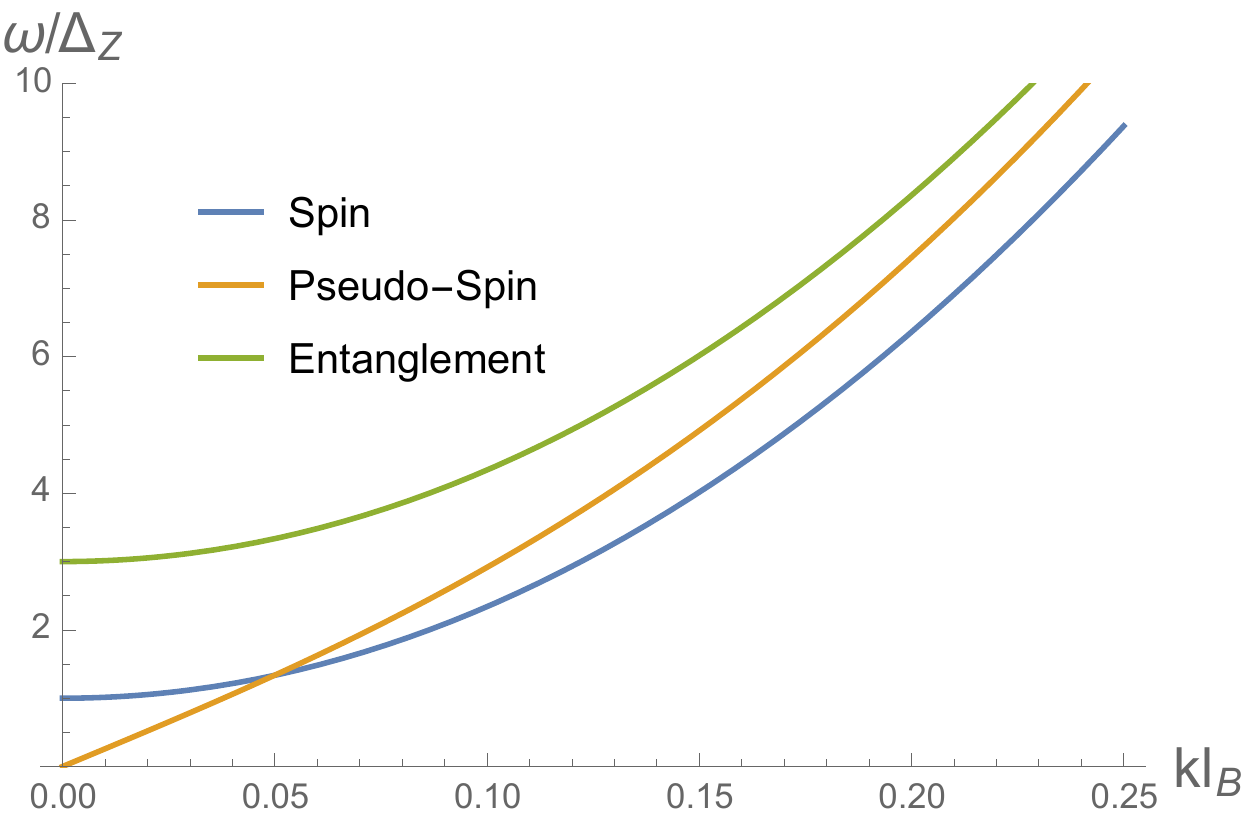} 
		\caption{Dispersion relation of the three modes in the KD phase for $u_z=2\Delta_Z$ and $u_\perp=-2\Delta_Z$. We observe that the pseudo-spin mode is gapless, linear at low momentum $|\mathbf{k}\ll k_0$ and becomes quadratic at higher momentum. the two other modes are gapped and quadratic.}
		\label{fig:DispKD}
	\end{center}
\end{figure}
The dispersion of the three modes is shown in Fig. \ref{fig:DispKD}. Analogously to the CDW case, the mode 1 corresponds to a spin mode, the mode 2 to a pseudo-spin mode and the mode 3 to an entanglement mode. 

The spin and entanglement modes are quite similar to the modes observed for the CDW phase, they are quadratic gapped modes with a gap proportional to the Zeeman coupling for the spin wave and a gap equal to $\Delta_Z-u_\perp$ for the entanglement wave, which corresponds to flipping both the spin and the pseudo-spin. This gap is always positive since in the KD phase, we have $u_\perp<\Delta_Z$. The space-time dependent spinor corresponding to these two mode has the same expression as (\ref{eq:space-time_spinors}) with the basis spinors given by Eqs. (\ref{eq:spinorsKD}). The pure spin wave has the spins of each sublattice oscillating at frequency $\omega_1$ with equal weight on both sublattices
\begin{align}
\mathbf{M_{S_A}}=\mathbf{M_{S_B}}=\frac{1}{2}\begin{pmatrix}
\pi_0\cos(\mathbf{k}\cdot\mathbf{r}-\omega_1t) \\ \pi_0\sin(\mathbf{k}\cdot\mathbf{r}-\omega_1t)\\	1-\frac{1}{2}\pi_0^2
\end{pmatrix} .
\label{eq:spin_waveKD}
\end{align}

Finally, the second mode looks different, it has a gapless linear dispersion at low-momentum $\mathbf{k}^2\ll u_z-u_\perp$ while we recover a quadratic dispersion relation at high momentum. The transition between these two regimes occurs at a momentum of $k_0=\sqrt{u_z-u_\perp}$. Similarly to the pseudo-spin mode in the CDW phase (\ref{eq:pseudo-spinCDW}), the energy $u_z-u_\perp$ corresponds to the energy necessary to bring one pseudo-spin out of the plane, namely there is a cost in out-of-plane anisotropic energy $u_z$ but a gain in in-plane anisotropic energy $u_\perp$. This energy is always positive in the KD region since $u_z>u_\perp$. Thereby, at low momentum, there is not enough energy to bring one pseudo-spin out of the plane. The model corresponds thus to an $XY$ model where the pseudo-spin is restricted to the equator of the Bloch sphere and this mode is analogous to the linearly dispersing superfluid mode in Helium and in bilayer 2DEGs\cite{Fertig1989,Moon1995,MacDonald2001}. Its gaplessness originates from the U(1) symmetry of the ground state : there is no cost in anisotropic energy cost for rotating a pseudo-spin in the $xy$ plane. When the energy is larger than $u_z-u_\perp$, there is now enough energy to bring the pseudo-spin out of the plane and we recover the usual quadratic dispersion relation associated with the fact that the two generators are now canonically conjugate.

\subsection{Anti-ferrimagnetic phase}

\label{sec:SWAFI}

The unitary matrix that tranforms the CDW spinors (\ref{eq:spinorsCDW}) into the entangled spinors of the AFI phase is given by
\begin{align}
	U_\text{AFI}=e^{i\frac{\alpha_1}{2}\sigma_x\mathbf{m}\cdot\bm{\tau}},
	\label{eq:UAFI}
\end{align}
where $\mathbf{m}=(\sin\beta,-\cos\beta,0)$ and $\alpha$ is given by Eq. (\ref{eq:alphaAFI}). The basis spinors of the AFI phase are
\begin{subequations}
	\begin{align}
	|\tilde{F}\rangle&=\cos\frac{\alpha_1}{2}|\mathbf{n}_z\rangle|\mathbf{s}_z\rangle+e^{i\beta}\sin\frac{\alpha_1}{2}|-\mathbf{n}_z\rangle|-\mathbf{s}_z\rangle  \label{eq:F_AFI} \\
	|\tilde{C}_1\rangle&=\cos\frac{\alpha_1}{2}|\mathbf{n}_z\rangle|-\mathbf{s}_z\rangle+e^{i\beta}\sin\frac{\alpha_1}{2}|-\mathbf{n}_z\rangle|\mathbf{s}_z\rangle \\
	|\tilde{C}_2\rangle&=-\sin\frac{\alpha_1}{2}e^{-i\beta}|\mathbf{n}_z\rangle|-\mathbf{s}_z\rangle+\cos\frac{\alpha_1}{2}|-\mathbf{n}_z\rangle|\mathbf{s}_z\rangle \\
	|\tilde{C}_3\rangle&=-\sin\frac{\alpha_1}{2}e^{-i\beta}|\mathbf{n}_z\rangle|\mathbf{s}_z\rangle+\cos\frac{\alpha_1}{2}|-\mathbf{n}_z\rangle|-\mathbf{s}_z\rangle. \label{eq:C3_AFI}
	\end{align}
	\label{eq:spinorsAFI}
\end{subequations}
We can see that the modes 1 and 2 involves the four basis spinors $|\mathbf{n}_z\rangle|\mathbf{s}_z\rangle$, $|-\mathbf{n}_z\rangle|-\mathbf{s}_z\rangle$, $|-\mathbf{n}_z\rangle|\mathbf{s}_z\rangle$ and $|\mathbf{n}_z\rangle|-\mathbf{s}_z\rangle$ and one cannot factor the spinors in order to have a definite spin or pseudo-spin mode. We find that these two modes are coupled and their dispersions are given by
\begin{align}
	\omega_{1,2}&=\pm\left(\frac{u_\perp}{2}-u_z\right)\cos\alpha_1 \nonumber \\
	&+\sqrt{2\pi\rho_s(\mathbf{k}l_B)^2[2\pi\rho_s(\mathbf{k}l_B)^2+2u_\perp]+\frac{u_\perp^2}{4}\cos^2\alpha_1} ,
\end{align}
which are both positive due to the gap term inside the square root. The gaps $\Delta_\alpha$ of the modes $\alpha=1$ and $\alpha=2$ are 
\begin{align}
	\Delta_1&=u_z\cos\alpha_1=\Delta_Z \\
	\Delta_2&=\left(u_\perp-u_z\right)\cos\alpha_1 .
\end{align}
\begin{figure}[t]
	\begin{center}
		\includegraphics[width=8cm]{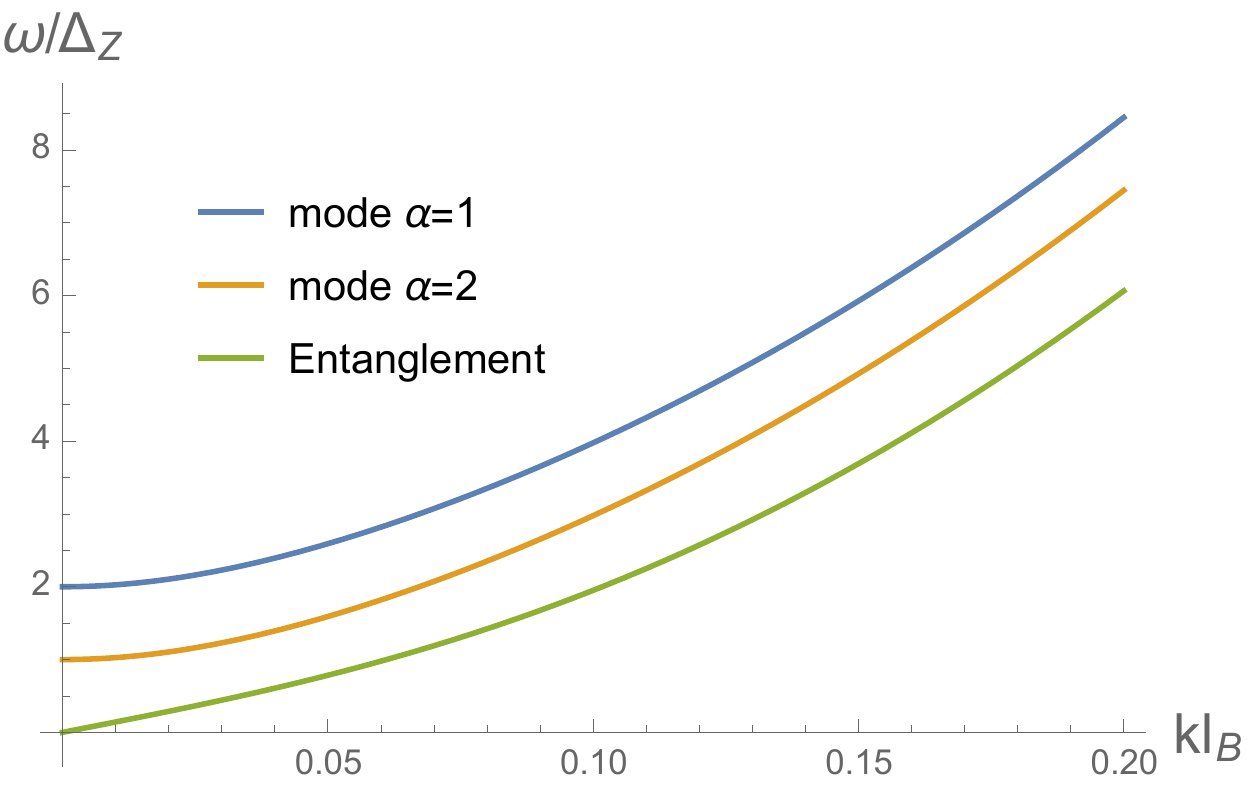} 
		\caption{Dispersion relation of the three modes in the AFI phase for $u_z=2\Delta_Z$ and $u_\perp=6\Delta_Z$. The modes $a=1$ and $a=2$ are coupled and form the $\alpha=1$ and $\alpha=2$ which are quadratically dispersing while the entanglement mode is linear and gapless.}
		\label{fig:DispAFI}
	\end{center}
\end{figure}
For $\alpha=0$, namely at the boundary with the CDW phase, the spinors (\ref{eq:spinorsAFI}) simplify to the CDW spinors and we recover the pseudo-spin mode with gap $u_\perp-u_z$ and the spin mode with gap $\Delta_Z=u_z$.

The dispersion for the entanglement mode $a=3$ is given by
\begin{align}
\omega_3(\mathbf{k})=\sqrt{2\pi\rho_s}|\mathbf{k}|l_B\sqrt{2\pi\rho_s(\mathbf{k}l_B)^2+u_z(1-\cos^2\alpha_1)}.
\end{align}
We can see that for $\cos\alpha_1=1$ ($\Delta_Z=2u_z$), namely at the transition with the CDW phase, we obtain a gapless quadratic dispersion. When $\Delta_Z<2u_z$, we have a linear dispersion at low momentum which transforms into a quadratic dispersion around momentum $k_0=\sqrt{2u_z(1-\cos^2\alpha_1)}$. This mode is analogous to the pseudo-spin mode in the KD phase. The linearity at low-momentum originates from the U(1) symmetry of the ground state associated with the parameter $\beta$ in Eqs. (\ref{eq:F_AFI}) and (\ref{eq:C3_AFI}). The spinors $|\tilde{F}\rangle$ and $|C_3\rangle$ are both in a superposition of the states $|\mathbf{n}_z\rangle|\mathbf{s}_z\rangle$ and $|-\mathbf{n}_z\rangle|-\mathbf{s}_z\rangle$ as shown in Fig. \ref{fig:BlochSphereAFI}. It costs thus no anisotropic energy to move the ground state (black arrow in Fig. \ref{fig:BlochSphereAFI}) around the parallel of the Bloch sphere at which lie both the black and red arrows. At higher momentum, there is enough energy to bring the entanglement mode out of this latitude and restore the symmetry between the $xy$ direction and the $z$ direction.

\subsection{Canted anti-ferromagnetic phase}

\begin{figure}[t]
	\begin{center}
		\includegraphics[width=8.8cm]{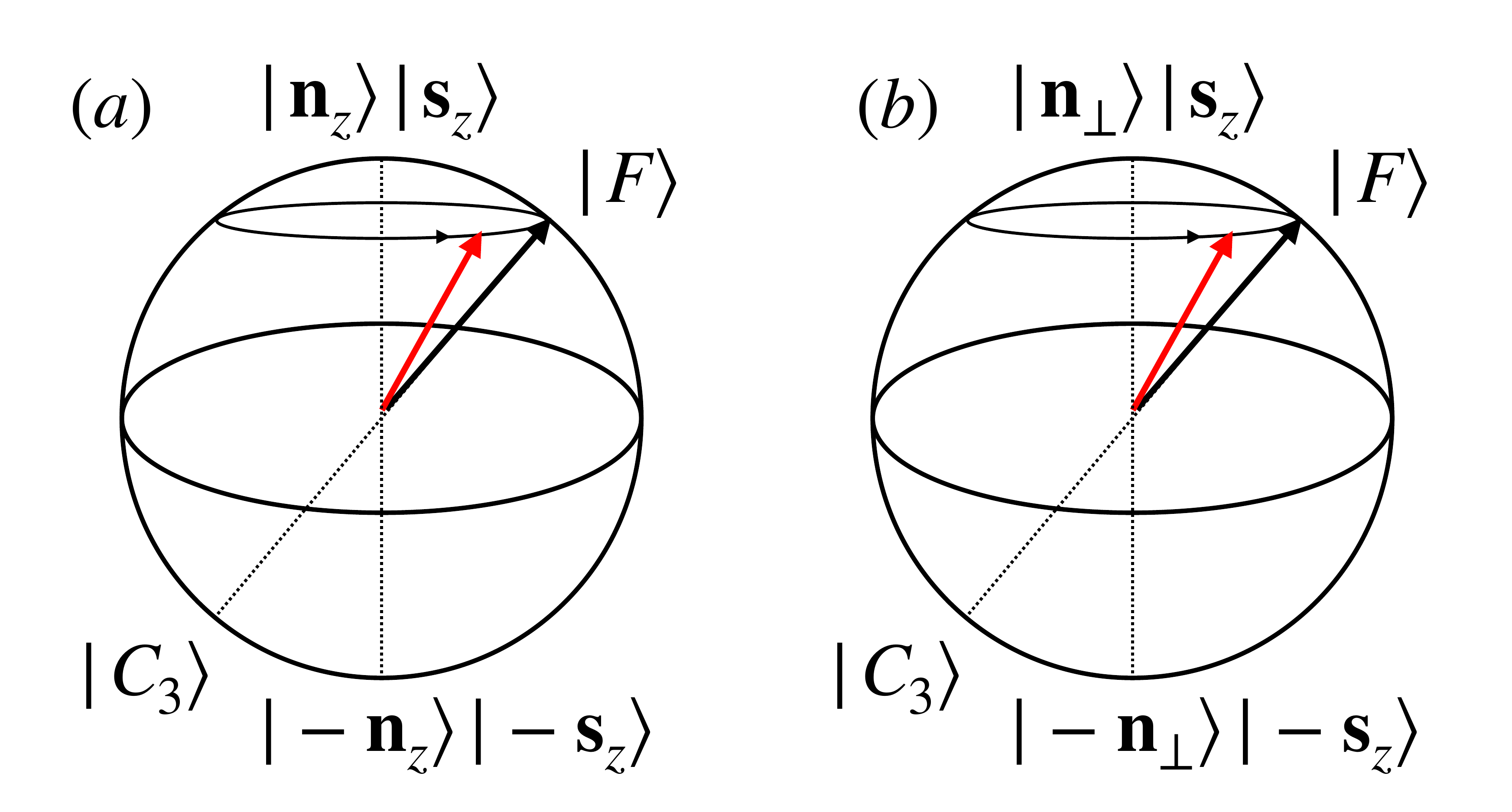} 
		\caption{Entanglement Bloch spheres corresponding to the entanglement mode in (a) the AFI phase and (b) the CAF phase. The spinor $|F\rangle$ indicated by the black arrow corresponds to the ground state, while the spinor $|C_3\rangle$ is located at opposite direction of the Bloch sphere. The ground states possesses a U(1) symmetry associated with the angle $\beta$ corresponding to the latitude indicated by the circle at the tip of the black and red arrows. At low-energy, the entanglement wave correponds to a small deviation at equi-latitude indicated by the red arrow.}
		\label{fig:BlochSphereAFI}
	\end{center}
\end{figure}

The unitary matrix that tranforms the CDW spinors (\ref{eq:spinorsCDW}) into the entangled spinors of the canted anti-ferromagnetic phase is the product of the matrices (\ref{eq:UKD}) and (\ref{eq:UAFI}) of the KD and AFI phase
\begin{align}
U_\text{CAF}=e^{i\frac{\pi}{4}\mathbf{n}\cdot\bm{\tau}}e^{i\frac{\alpha_2}{2}\sigma_x\mathbf{m}\cdot\bm{\tau}},
\end{align}
where $\mathbf{n}=(\sin\varphi,-\cos\varphi,0)$, $\mathbf{m}=(\sin\beta,-\cos\beta,0)$ and $\alpha_2$ is given by Eq. (\ref{eq:alphaCAF}).
\begin{figure}[t]
	\begin{center}
		\includegraphics[width=8cm]{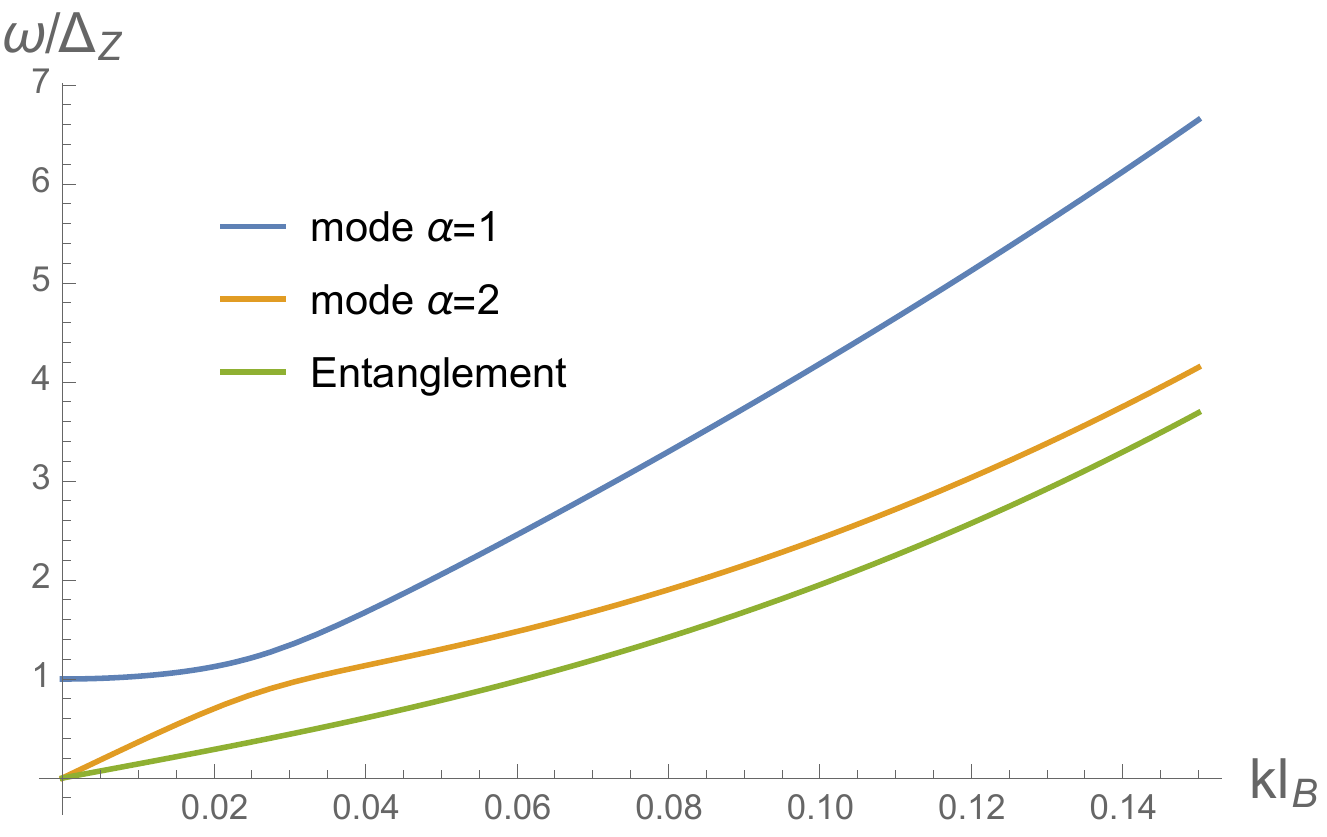} 
		\caption{Dispersion relation of the three modes in the CAF region for $u_z=12\Delta_Z$ and $u_\perp=2\Delta_Z$. We observe two gapless modes : the entanglement mode and the mode $\alpha=2$ which originates from the gapless mode of the KD region. }
		\label{fig:DispCAF12}
	\end{center}
\end{figure}

The basis spinors of the AFI phase are
\begin{subequations}
	\begin{align}
	|\tilde{F}\rangle&=\cos\frac{\alpha_2}{2}|\mathbf{n}_\perp\rangle|\mathbf{s}_z\rangle+e^{i\beta}\sin\frac{\alpha_2}{2}|-\mathbf{n}_\perp\rangle|-\mathbf{s}_z\rangle\label{eq:F_CAF} \\
	|\tilde{C}_1\rangle&=\cos\frac{\alpha_2}{2}|\mathbf{n}_\perp\rangle|-\mathbf{s}_z\rangle+e^{i\beta}\sin\frac{\alpha_2}{2}|-\mathbf{n}_\perp\rangle|\mathbf{s}_z\rangle \\
	|\tilde{C}_2\rangle&=-\sin\frac{\alpha_2}{2}e^{-i\beta}|\mathbf{n}_\perp\rangle|-\mathbf{s}_z\rangle+\cos\frac{\alpha_2}{2}|-\mathbf{n}_\perp\rangle|\mathbf{s}_z\rangle \\
	|\tilde{C}_3\rangle&=-\sin\frac{\alpha_2}{2}e^{-i\beta}|\mathbf{n}_\perp\rangle|\mathbf{s}_z\rangle+\cos\frac{\alpha_2}{2}|-\mathbf{n}_\perp\rangle|-\mathbf{s}_z\rangle  \label{eq:C3_CAF}
	\end{align}
\end{subequations}

The modes $a=1$ and $a=2$ are also coupled and we don't present their explicit expression here since it is too lengthy. We find the corresponding gaps
\begin{align}
\Delta_1&=\Delta_Z \\
\Delta_2&=0,
\end{align}
such that one mode is gapless with a linear dispersion relation at low-energy as can be seen in Fig. \ref{fig:DispCAF12} and one mode has a pure Zeeman gap. We can see that the modes $\alpha=1$ and $\alpha=2$ originate from an anti-crossing around momentum $|\mathbf{k}|l_B\approx 0.03$ between a linear mode and a gapped quadratic mode, which are the descendants of the spin and the gapless pseudo-spin modes of the KD phase. The mode 1 becomes quadratic at higher energy.

Once again, the mode 3 is decoupled from the others, and corresponds thus to an entanglement mode with dispersion
\begin{align}
\omega_3(\mathbf{k})=\sqrt{2\pi\rho_s}|\mathbf{k}|l_B\sqrt{2\pi\rho_s(\mathbf{k}l_B)^2+u_\perp(1-\cos^2\alpha_2)}.
\end{align}
This mode is the analog of the entanglement mode in the AFI phase except that we are in the basis $\{|\mathbf{n}_\perp\rangle|\mathbf{s}_z\rangle,|-\mathbf{n}_\perp\rangle|-\mathbf{s}_z\rangle\}$ as shown in Fig. \ref{fig:BlochSphereAFI}. The gaplessness and linearity originates also from the U(1) symmetry associated with the angle $\beta$ in Eqs. (\ref{eq:F_CAF}) and (\ref{eq:C3_CAF}).

\section{Conclusion}

\begin{figure}[t]
	\begin{center}
		\includegraphics[width=8cm]{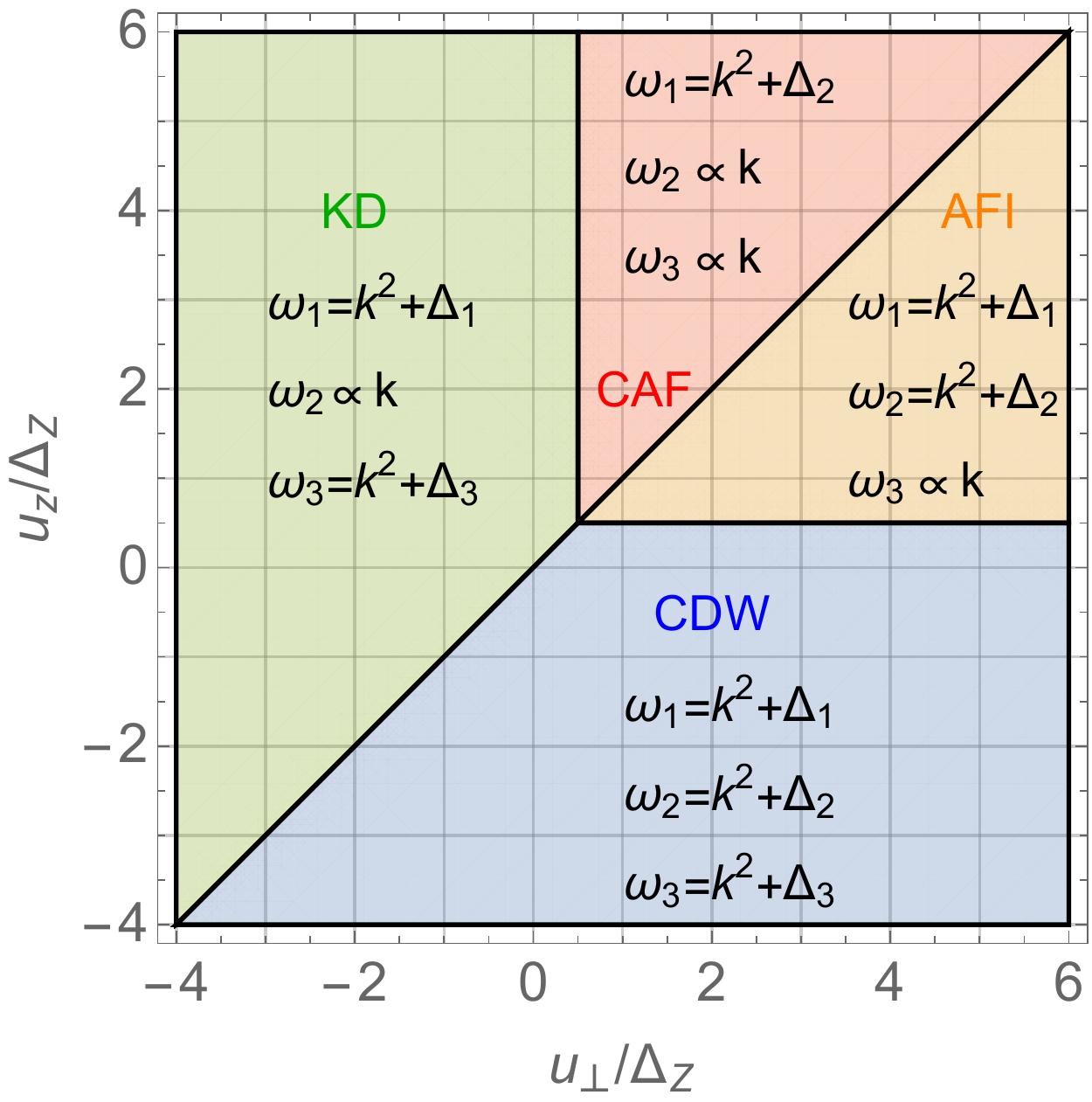} 
		\caption{Summary of the low-energy dispersion relation in the four phases. The indices 1,2 and 3 refer to the spin, pseudo-spin and entanglement modes respectively, except in the CAF and AFI region where the spin and pseudo-spin modes are coupled. In the schematic expression of the dispersion relations, we have set $\rho_s/\rho_0=2\pi\rho_sl_B^2\equiv 1$. In the CDW region, the three modes are gapped. In the KD region, there are two gapped modes and one gapless linear modes, the pseudo-spin mode. In the AFI region, the entanglement mode is gapless while the two other modes are gapped. Finally, in the CAF region, there are two gapless modes, the entanglement and the coupled mode $\alpha=2$, the descendant of the pseudo-spin mode.}
		\label{fig:Summary}
	\end{center}
\end{figure}

To conclude, we have presented the dispersions of the different types of spin waves, namely pure spin, valley, and entanglement waves, in graphene at filling factor $\nu=\pm1$. We have considered the four different possible ground states presented by Lian \textit{et al}\cite{Lian2017} based on the anisotropic terms $u_\perp$ and $u_z$ originally introduced by Kharitonov\cite{Kharitonov2012}. We have introduced a non-linear sigma model based on a Lagrangian formalism which describes the long wavelength space-time dependent spin-valley rotations. The presence of small explicit symmetry-breaking terms generally opens a gap in the dispersion relation of the different types of spin waves. However, we have found that in each phase, except in the CDW region, there remain one or two gapless modes with a linear dispersion relation at low momentum. The fact that these modes remain gapless originates from a residual symmetry of the ground state, which is present even when the symmetry breaking terms are introduced. These modes recover a quadratic dispersion relation at higher energies when the symmetry between the different directions of oscillation is restored. The summary of our findings for the presence or absence of a gap for the three modes in each region is presented in Fig. \ref{fig:Summary}.

Our study, along with the expression for the gaps at $\nu=0$ for the KD and CAF phase presented in Ref. [\onlinecite{Wu2014a}] opens the way to an analysis of the scattering of spin waves at interfaces between regions with different filling factor taking into account the different types of spin wave (spin, pseudo-spin or entanglement). Depending on the steepness of the scattering region, we expect a different scattering process and emit the possibility that one wave type in the $\nu=\pm1$ region might be changed in the scattering process, or be in a superposition of different types, since the type of spin waves are different at $\nu=0$. The scattering mechanism should also depend on the phase the region at $\nu=0$ is in.

\begin{acknowledgements}
We would like to thank Alexandre Assouline, Preden Roulleau, Rebeca Ribeiro Palau, and Fran\c cois Parmentier for stimulating discussions. We acknowledge financial support from Agence Nationale de la Recherche (ANR project ``GraphSkyrm'') under Grant No. ANR-17-CE30-0029.
\end{acknowledgements}

\bibliographystyle{apsrev4-1}
\bibliography{library}

\end{document}